\documentclass[aps,twocolumn,showpacs,preprintnumbers,amsmath,amssymb,nofootinbib,superscriptaddress,showkeys]{revtex4}

\usepackage{epsfig}
\usepackage{graphicx}
\usepackage{mathptmx}      
\usepackage{latexsym}

\def\slashchar#1{\setbox0=\hbox{$#1$}
   \dimen0=\wd0 \setbox1=\hbox{/} \dimen1=\wd1
   \ifdim\dimen0>\dimen1 \rlap{\hbox to \dimen0{\hfil/\hfil}} #1
   \else  \rlap{\hbox to \dimen1{\hfil$#1$\hfil}} / \fi}

\begin{document}

\title{Systematics of radial and angular-momentum Regge trajectories of light non-strange $q\bar{q}$-states%
\footnote{Supported by
MICINN of Spain (FPA2010-16802, FPA2010-16696, FIS2011-24149)
and Consolider-Ingenio   2010 Programme CPAN
(CSD2007-00042), by Junta de Andaluc\'{\i}a
(FQM 101, FQM 437, FQM225 and  FQM022) and by the Polish Science and Higher Education, grant N~N202 263438, and National Science
Centre, grant DEC-2011/01/D/ST2/00772.}}

\author{Pere Masjuan} \email{masjuan@ugr.es}
\affiliation{Departamento de F\'{i}sica Te\'{o}rica y del Cosmos and CAFPE,
Universidad de Granada, E-18071 Granada, Spain}

\author{Enrique Ruiz Arriola}\email{earriola@ugr.es}
\affiliation{Departamento de
  F\'{\i}sica At\'omica, Molecular y Nuclear
and Instituto Carlos I de F{\'\i}sica Te\'orica y Computacional \\
Universidad de Granada, E-18071 Granada, Spain.}

\author{Wojciech Broniowski} \email{Wojciech.Broniowski@ifj.edu.pl}
\affiliation{The H. Niewodnicza\'nski Institute of Nuclear Physics,
  PL-31342~Krak\'ow, Poland}
\affiliation{Institute of Physics, Jan Kochanowski University, PL-25406~Kielce, Poland}

\date{\today}

\begin{abstract}
We reanalyze the radial ($n$) and angular-momentum ($J$) Regge
trajectories for all light-quark states with baryon number zero listed
in the 2011 edition of the Particle Data Tables.  The parameters of
the trajectories are obtained with linear regression, with weight of
each resonance inversely proportional to its half-width squared,
$(\Gamma/2)^2$. That way we are side-stepping possible
channel-dependent and model-dependent extractions of the resonance
parameters and are able to undertake an error analysis. The method
complies to the fact that the pole position of the resonance is
typically shifted from channel-dependent extractions by $\sim
\Gamma/2$. This is also a feature of the large-$N_c$ limit of QCD,
where the masses change by $~\Gamma/2$ when evolving from $N_c=3$ to
$N_c=\infty$. Our value for the slope of the radial Regge trajectories
is $a=1.35(4) {\rm GeV}^2$. We discuss the fundamental issue whether the masses of the light-quark non-strange
states fit into a universal pattern $M_{nJ}^2 = a (n+J) +b$, as suggested by Afonin, and
also predicted by some holographic models.
Our joint linear-regression analysis in the $(n,J,M^2)$ Regge planes
indicates, at a statistically significant level of 4.5 standard deviations, that the slopes of the radial Regge trajectories are
larger from the angular-momentum slopes. Thus no strict universality of slopes occurs in the light non-strange meson spectra.
\end{abstract}



\maketitle

\section{Introduction}

The study of regularities in the hadronic spectrum has been a
recurrent subject in the quark model~\cite{Klempt:2007cp}, as it allows
not only to check our current understanding of strong interactions, but
also to predict possible missing states. In the case of light-quark mesons, which
is the subject of the present study, the quark-hadron
duality~\cite{Shifman:2001qm} implies QCD constraints based on the
Operator Product Expansion of a two-point correlation function
with some given mesonic quantum numbers (say $J$). In particular,
\begin{eqnarray}
 f_{nJ}^2 / (d M_{nJ}^2 / dn) \to {\rm const},
\label{eq:Regge-asymp}
\end{eqnarray}
where $M_{nJ}$ is the n-th mass of the meson and $f_{nJ}$ the
corresponding vacuum decay amplitude. More than a decade ago Anisovich,
Anisovich, and Sarantsev~\cite{Anisovich:2000kxa} suggested that mesons
could be grouped into {\it radial} Regge trajectories of the form
\begin{equation}
M_{n}^2\,=\,M^2_0 + n \mu^2\, ,
\label{Regge}
\end{equation}
where $M_0$ is the mass of the lowest-lying meson on each
corresponding trajectory and $\mu^2$ is the slope
parameter.  According to Ref.~\cite{Anisovich:2000kxa}, the slope is
approximately the same for {\it all} the trajectories considered:
$\mu^2=1.25(15)$~GeV$^2$.  The uncertainly was estimated based
on the spread of the different results for each meson family.  In
addition, some missing states predicted from Eq.~(\ref{Regge}) have
indeed been confirmed and included in the latest edition of the
Particle Data Group (PDG) tables~\cite{Nakamura:2010zzi}. Furthermore,
Ref.~\cite{Anisovich:2000kxa} also analyzed the venerable
angular-momentum Regge trajectories~\cite{Chew:1961ev} (for a review
see, e.g., \cite{Collins:1971ff}), which motivated the original
(rubber) string models~\cite{Susskind:1970qz} (for a review see,
e.g.,~\cite{Jacob:1974,Frampton:1986wv}).  Moreover, the large
degeneracy~\cite{Huang:1970iq} of the daughter Regge trajectories is
capable of producing the Hagedorn growth of the hadronic
spectra~\cite{Hagedorn:1965st,Huang:1970iq} (see
Refs.~\cite{Broniowski:2000bj,Broniowski:2004yh} for a recent
reanalysis).

In a remarkable paper Afonin~\cite{Afonin:2006wt} (see
also~\cite{Glozman:2007ek}) analyzed jointly the radial and
angular-momentum trajectories and argued that they merge into a single
pattern
\begin{equation}
M^2(n,J)\,=\,a(n+J) + c\, , \label{eq:afo}
\end{equation}
unveiling a kind of hydrogen-like accidental degeneracy, with a
harmonic oscillator mass-squared spectrum. All these phenomenological
findings provide some confidence on the string picture of hadrons,
where the square of the mass is the fundamental dynamical
quantity. Together with the QCD short-distance constraint
of Eq.~(\ref{eq:Regge-asymp}) we may then infer that mesonic vacuum decay
amplitudes tend to a constant in the upper part of the spectrum.

Regardless of the success of the radial Regge trajectories, it is
important to note that the resonance parameters, such as
mass, width, or coupling constants, depend on the definitions and are
sensitive to the background, i.e., to the particular process used to
extract the resonance from the experimental data.  This poses the
relevant question of what the precise meaning of Eq.~(\ref{Regge}) is,
and, moreover, in what sense is QCD compatible with such an
analysis. In the present work we reanalyze this problem, carrying out
global linear regression fits with the uncertainty of the resonance
position proportional to its width, $\Gamma$. Specifically, we use
weights inversely proportional to the square of the resonance
half-width.  The approach is consistent with the fact that the pole
position of the resonance is typically shifted from channel-dependent
extractions by about $\Gamma/2$. Also, within the large-$N_c$
QCD~\cite{'tHooft:1973jz,Witten:1979kh} (see e.g. \cite{Pich:2002xy}
for a review), where the strong coupling constant is assumed to scale
as $ g \sim 1/ \sqrt{N_c}$, the meson masses change by $~\Gamma/2$
when evolved from $N_c=3$ to $N_c=\infty$, as has been exploited intensely in
Refs.~\cite{Harada:2003em,Pelaez:2003dy,Pelaez:2006nj,Nieves:2009kh,Nieves:2009ez,Nieves:2011gb,Nebreda:2011cp,Guo:2011pa}.

We note that within the AdS/CFT proposal (for a review see,
e.g.,~\cite{Erdmenger:2007cm}) there have been attempts to formulate
holographic models (the so-called soft-wall models) with linear
confinement~\cite{Karch:2006pv} and, likewise, their light-cone
relatives~\cite{deTeramond:2008ht}, complying to the ansatz of Eq.~(\ref{Regge}).
We recall that all these AdS/CFT inspired models
are claimed to operate for large t'Hooft couplings, i.e., $ g \sim 1/N_c$.

As we will elaborate in detail, our main finding, after considering
the resonance width uncertainties, is to confirm the result of
Ref.~\cite{Anisovich:2000kxa} with the updated data,
as we find $\mu^2=1.35(4)~{\rm GeV}^2$.
On the other hand, our analysis in the $(n,J,M^2)$ Regge planes shows that at a statistically significant level of 4.5 standard deviations
the slopes of the radial Regge trajectories are
larger from the slopes of the angular-momentum trajectories. Therefore no strict universality of slopes occurs in the light non-strange meson sector.

The plan of the paper is as follows. In Section~\ref{sec:res-largeN}
we motivate our choice for the weight in the linear regression analysis.

In Section~\ref{sec:radial-Regge} we discuss in detail, through the use of
the present PDG tables, how the different states are grouped into the radial
Regge trajectories. Whenever possible, we try to keep the
successful choice of Ref.~\cite{Anisovich:2000kxa} taking into account
the assumed uncertainties. In Section~\ref{sec:radial-Regge2} on we enlarge
the choice of Ref.~\cite{Anisovich:2000kxa} to complete all the light
unflavored states collected in the PDG. The update of the angular-momentum Regge
trajectories is considered in Section~\ref{sec:J-Regge}.  In
Section~\ref{sec:merge} we discuss, as originally suggested by Afonin,
simultaneously the radial and angular-momentum trajectories. Finally, in
Section~\ref{sec:concl} we summarize our results and draw our main
conclusions.

Throughout this work we use the up-to-date edition of the PDG
tables~\cite{Nakamura:2010zzi}.  The symbol $q$ stand for the light quarks, $u$ or $d$.

\section{Uncertainties of resonance positions \label{sec:res-largeN}}

As already mentioned in the Introduction, in order to properly size
the meaning of the radial Regge trajectories for resonant states it is
important to review the well known features of the quantum mechanical
decay process relevant to our discussion.  The rigorous
quantum-mechanical definition of a resonance with given quantum
numbers corresponds to a pole in the second Riemann sheet in the
(analytically continued) partial-wave amplitude of the considered
scattering channel~\cite{Martin:102663}. This definition becomes
independent on the background, whereas the corresponding residue
provides the amplitude to produce that resonance in the given process.

However, although quoting the complex pole and the complex residue
would be superior and highly desirable, for practical reasons this is
not what one typically finds in the PDG
tables~\cite{Nakamura:2010zzi}, with very few exceptions.  As a matter
of fact, several definitions besides the pole in the second Riemann
sheet are employed, such as a pole in the $K$-matrix, the Breit-Wigner
resonance, the location of a maximum in the speed plot, time delay,
etc.  (see, e.g.,~\cite{Suzuki:2008rp,Workman:2008iv}).

A resonance may be interpreted as a superposition of states with a
given mass distribution on the real axis, approximately spanning the
$M \pm \Gamma/2$ interval. Of course, the shape of the distribution
depends on the particular process in which the resonance is produced,
and thus on the background. Clearly, while all the definitions
converge for narrow resonances, even for broad states we expect the
masses obtained from various methods to be compatible within their
corresponding $M \pm \Gamma/2$ intervals. As stated above, the values
listed by the PDG for a given resonance correspond to different
choices of the definition and/or production processes, but mostly the
results are compatible within the estimated width differences.  This
clearly provides an upper bound on the uncertainty of the resonance
position for different resonance parameter definitions.  For
shortness, we refer to this mass uncertainty estimate of the resonance
mass as the {\it half-width rule}~\footnote{Of course, the width
  itself has an uncertainty which may eventually enlarge the global
  indetermination in the resonance mass.}.

\begin{figure}
\centering
  \includegraphics[width=0.45\textwidth]{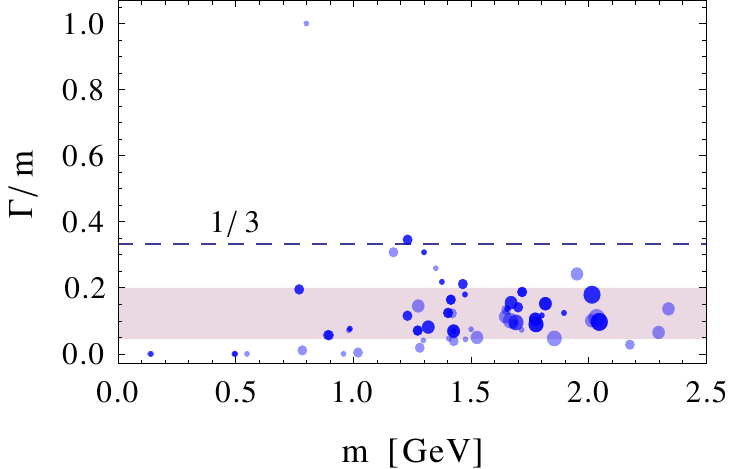}\\
  \caption{The ratio of width to mass for the light-quark meson
    states. The surface of each point is proportional to the $(2J +
    1)$ spin degeneracy, while the intensity is proportional to the
    isospin degeneracy $(2I + 1)$. The band correspond to the average
    $\pm$ standard deviation bounds, $\Gamma/M =
    0.12(8)$. \label{fig:ratio}}
\end{figure}

Quite remarkably, there is a QCD scenario where the half-width rule
estimate becomes parametrically small for {\it all} the resonances in
the mesonic spectrum. In the large-$N_c$ limit of
QCD~\cite{'tHooft:1973jz,Witten:1979kh} (see e.g. \cite{Pich:2002xy}
for a review using effective Lagrangians) mesons become stable,
i.e., their masses are $M = {\cal O} (N_c^0)$, while their widths are
suppressed, $\Gamma= {\cal O}(1/N_c)$, such that the ratio $\Gamma/M =
{\cal O}(1/N_c)$. This expectation of the large-$N_c$ limit seems to
be fulfilled very well in the real $N_c=3$ world, since one finds for the light-quark mesons an
average value $ \Gamma/M = 0.12(8)$ (to be compared with a
rule-of-thumb $1/N_c=0.33$ for $N_c=3$)~\cite{Arriola:2011en}.
This feature is visualized in Fig.~\ref{fig:ratio}.
Of course, there are exceptions to
this average ratio, but they are scarce within the given confidence
interval. In fact, only just one state ($\sigma=f_0(600)$) goes over the
$1/3$-value~\cite{Arriola:2011en}.

A complementary way of connecting parametrically the mass shift and
the decay width is as follows.  One starts with the leading-$N_c$
resonance Lagrangian~\cite{Pich:2002xy}, recalling that the three- and
higher n-mesonic interactions are ${\cal
  O}(N_c^{1-n/2})$~\cite{'tHooft:1973jz,Witten:1979kh}. Thus, the mass
shift is computed as a loop integral via the self-energy whose
imaginary part corresponds to the decay width of the particles inside
the loop according to the Cutkosky rules. This argument makes it clear
that the $1/N_c$ scaling of the mass-shift and the width are exactly
the same and bound by ${\cal O}(1/N_c)$, although the numerical values
of the two quantities may not coincide exactly. The point of this
discussion is that if we take the leading-$1/N_c$ resonance mass, its
{\it systematic} uncertainty
is parametrically indistinguishable from the decay width, since they are
the real and imaginary parts of the self-energy, respectively. As
pointed out in Ref.~\cite{Masjuan:2007ay}, the role of the mass-shift
is crucial when determining the properties of two-point correlator
functions.

Within this framework, the half-width rule has been used
recently~\cite{RuizArriola:2010fj,Arriola:2011en} for the case of the
scalar and pseudoscalar mesons with rather interesting results
regarding the identification of glueball states and chiral symmetry
doublets. Here we extend these ideas to the rest of the light-quark
meson spectrum.
Specifically, to incorporate the half-width rule in practice,
we take~\footnote{There is an alternative fit with $\chi^2 = \sum_n
  \left(\frac{M_{n}-M_{n, {\rm exp}}}{\Gamma_{n}/2} \right)^2 $ which
  does not alter much in the results. Actually, both $\chi^2$
  functions are particular examples of the more general
  maximum-likelihood method, where the resonance production profile is
  assumed to be Gaussian. For a discussion on other profiles, in
  particular for the ubiquitous Breit-Wigner shape, see
  Appendix~\ref{sec:app} for details.}
\begin{eqnarray}
\chi^2 = \sum_n \left(\frac{M_{n}^2-M_{n, {\rm exp}}^2}{\Gamma_{n} M_n} \right)^2 \, ,
\label{eq:chi2}
\end{eqnarray}
for the linear regression fit, where the radial Regge formula,
Eq.~(\ref{Regge}), is used as the model. Note that in doing so, we are
just saying that Eq.~(\ref{Regge}) is fulfilled within the uncertainty
$M_n^2 = \mu^2 n + M_0^2 \pm \Gamma_n M_n$.  Moreover, we will stay
within the linear ansatz as the half-width rule yields
insensitivity to small non-linearities as analyzed e.g. in
Ref.~\cite{Afonin:2004yb} for n-trajectories or in
Ref.~\cite{Tang:2000tb} for J-trajectories.

\section{Radial Regge trajectories \label{sec:radial-Regge}}

The construction of a meson Regge trajectory requires a choice on the
possible meson assignments.  The analysis of the radial Regge
trajectories we are carrying out consists of two stages: The first one
reanalyzes the results of Ref.~\cite{Anisovich:2000kxa} with the
inclusion of more states from the updated PDG
tables~\cite{Nakamura:2010zzi}, while from
Sec.~\ref{sec:radial-Regge2} on we deal with meson families not
considered in Ref.~\cite{Anisovich:2000kxa}. To facilitate the comparison,
we follow as close as possible the presentation of Ref.~\cite{Anisovich:2000kxa}.

We motivate our selections with rather detailed discussions. The reader interested in
the results only may jump to Sec.~\ref{sec:summary-radial}.

In all our $M^2$-plots we take, in line with the half-width rule, the error to be given by $\Delta M^2=\pm \Gamma M$.

\subsection{$a_1(11^{++})$ and $a_3(13^{++})$}

\begin{figure}
\centering
\includegraphics[width=0.45\textwidth]{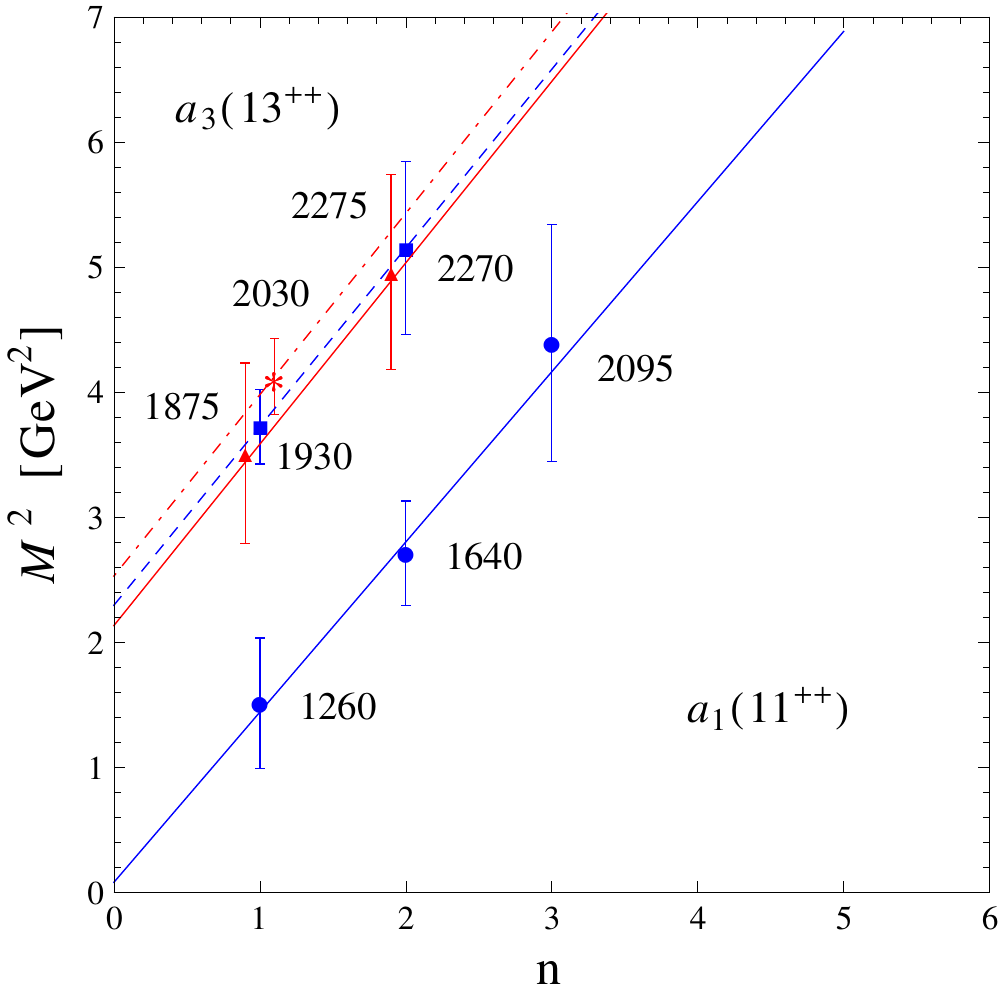}
\caption{(color online) The $(n,M^2)$ plots for the states $a_1(1^{++})$ (lower solid and dashed lines) and $a_1(3^{++})$ (upper solid and dot-dashed lines). Error bars correspond to take $\Delta M^2=\pm \Gamma M$. \label{a1plot}}
\end{figure}

Compared to Ref.~\cite{Anisovich:2000kxa}, we consider four different
trajectories: two for the $a_1(11^{++})$ states (lower solid and
dashed lines in Fig.~\ref{a1plot}) and two for the $a_3(13^{++})$
states (upper solid and dot-dashed lines in Fig.~\ref{a1plot}).  The
first trajectory for the $a_1(11^{++})$ states contains $a_1(1260)$,
$a_1(1640)$, and $a_1(2095)$. The $a_1(2340)$ state (now called
$a_1(2270)$), assumed in Ref.~\cite{Anisovich:2000kxa} to belong to
this trajectory, is now used in the daughter trajectory for the
$a_1(11^{++})$, together with a new state not considered in
Ref.~\cite{Anisovich:2000kxa}, the $a_1(1930)$.  The linear fit to the
first trajectory for the $a_1(11^{++})$ states yields $\mu^2=1.36(49){\rm GeV}^2$
with $\chi^2/{\rm DOF} = 0.12$ (lower-solid line in
Fig.~\ref{a1plot}).  For the second trajectory $\mu^2=1.43(73)~{\rm
  GeV^2}$ (dashed line on Fig.~\ref{a1plot}).  This trajectory has only
two states and it will not be considered for the final compilation.

In the case the $a_3(13^{++})$ trajectories, the first one contains
the new $a_3(1875)$ state together with the $a_3(2275)$, yielding
$\mu^2=1.5(1.1)~{\rm GeV^2}$ (upper solid line in
Fig.~\ref{a1plot}) and the second trajectory contains only the
$a_3(2030)$ state (dot-dashed line parallel to the $a_3(13^{++})$
trajectory).

\subsection{$\eta (00^{-+})$ and $\eta_2 (02^{-+})$}

\begin{figure}
\centering
  \includegraphics[width=0.45\textwidth]{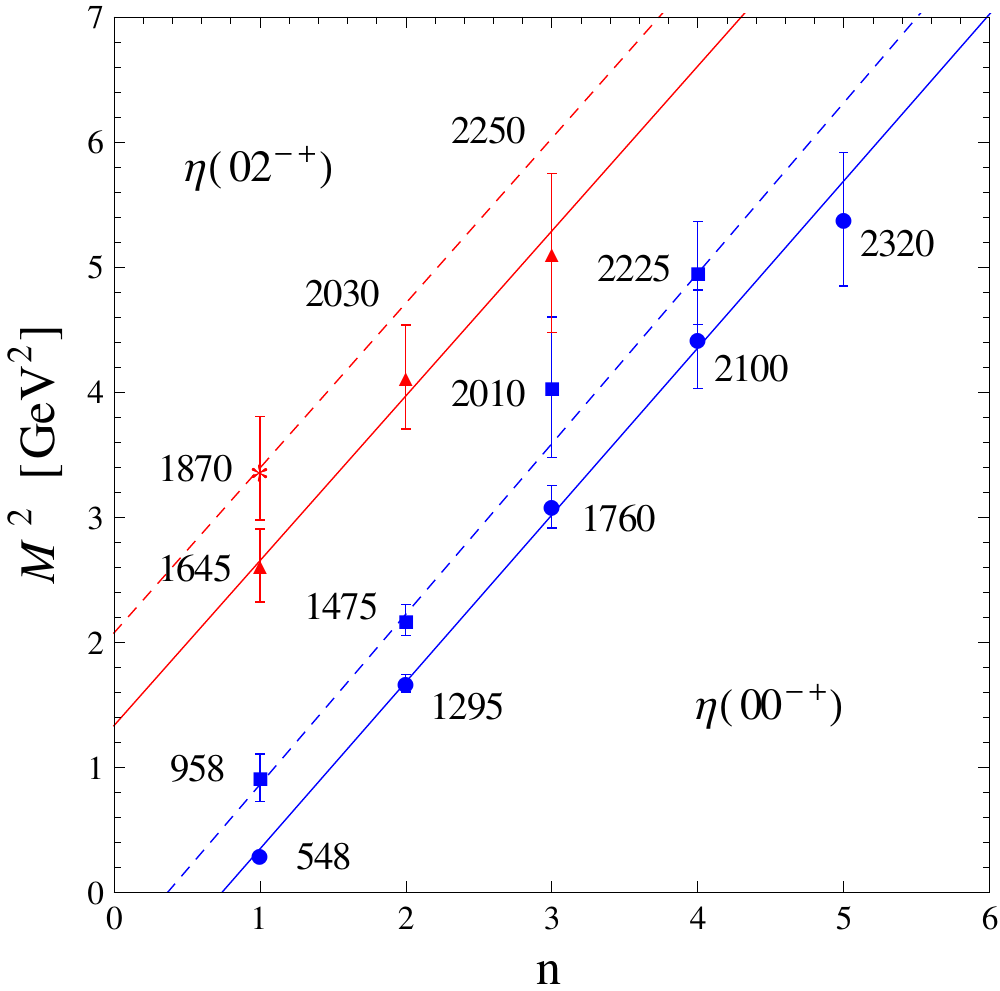}
  \caption{(color online) The $(n,M^2)$ plots for the $\eta (0 0^{-+})$ (lower solid and dashed lines) and $\eta_2 (0 2^{-+})$ (upper solid and dashed lines) trajectories. Error bars correspond to take $\Delta M^2=\pm \Gamma M$. \label{etaplot} }
\end{figure}

Figure~\ref{etaplot} shows the $\eta(00^{-+})$ and $\eta_2(02^{-+})$
states where, due to two independent flavor components $q\bar{q}$ and
$s\bar{s}$, both yield two trajectories.  The $\eta
(00^{-+})_{q\bar{q}}$ contains five states: $\eta(548)$, $\eta(1295)$,
$\eta(1760)$, $\eta(2100)$, and $\eta(2320)$ (lower solid line in
Fig.~\ref{etaplot}), where the first state, $\eta(548)$, is not used
in the linear fit. In Ref.~\cite{Anisovich:2000kxa} the state
$\eta(2100)$ was predicted, while nothing was said about the
$\eta(2320)$. Both states, now listed by the PDG, are incorporated in
our study.  The fit yields $\mu^2=1.33(11)~{\rm GeV}^2$ with $\chi^2/{\rm DOF}=0.26$.

The $\eta(00^{-+})_{s\bar{s}}$ trajectory with four states:
$\eta(958)$, $\eta(1475)$, $\eta(2010)$, and $\eta(2225)$, yields
$\mu^2=1.36(14)~{\rm GeV}^2$ with $\chi^2/{\rm DOF}=0.44$. The $\eta(2010)$ was
predicted in Ref.~\cite{Anisovich:2000kxa} under the name $\eta(1900)$
and now is listed in the PDG Tables.

In Ref.~\cite{Anisovich:2000kxa} only one state with mass near
$1440$~MeV was considered. Now it is well established that in this
energy region there are two different $\eta$ states, the $\eta(1405)$
and the $\eta(1475)$.  The first one, however, is not unambiguously
located and it is considered to be a glueball (see the mini-review
about this state on the PDG Tables), therefore we exclude it from our
fitting procedure. The second state is included in the $s\bar{s}$
trajectory.

The $\eta (02^{-+})_{q\bar{q}}$ trajectory yields $\mu^2=1.32(32)~{\rm GeV}^2$,
with $\chi^2/{\rm DOF}=0.22$. This trajectory contains
$\eta_2(1645)$, $\eta_2(2030)$, and $\eta_2(2250)$.  The $\eta
(02^{-+})_{s\bar{s}}$ trajectory, which contains only one
$\eta_2(1870)$ state, is drawn parallel to the non-strange case.

\subsection{$\rho_1(11^{--})$ and $\rho_3(13^{--})$}

\begin{figure}
\centering
  \includegraphics[width=0.45\textwidth]{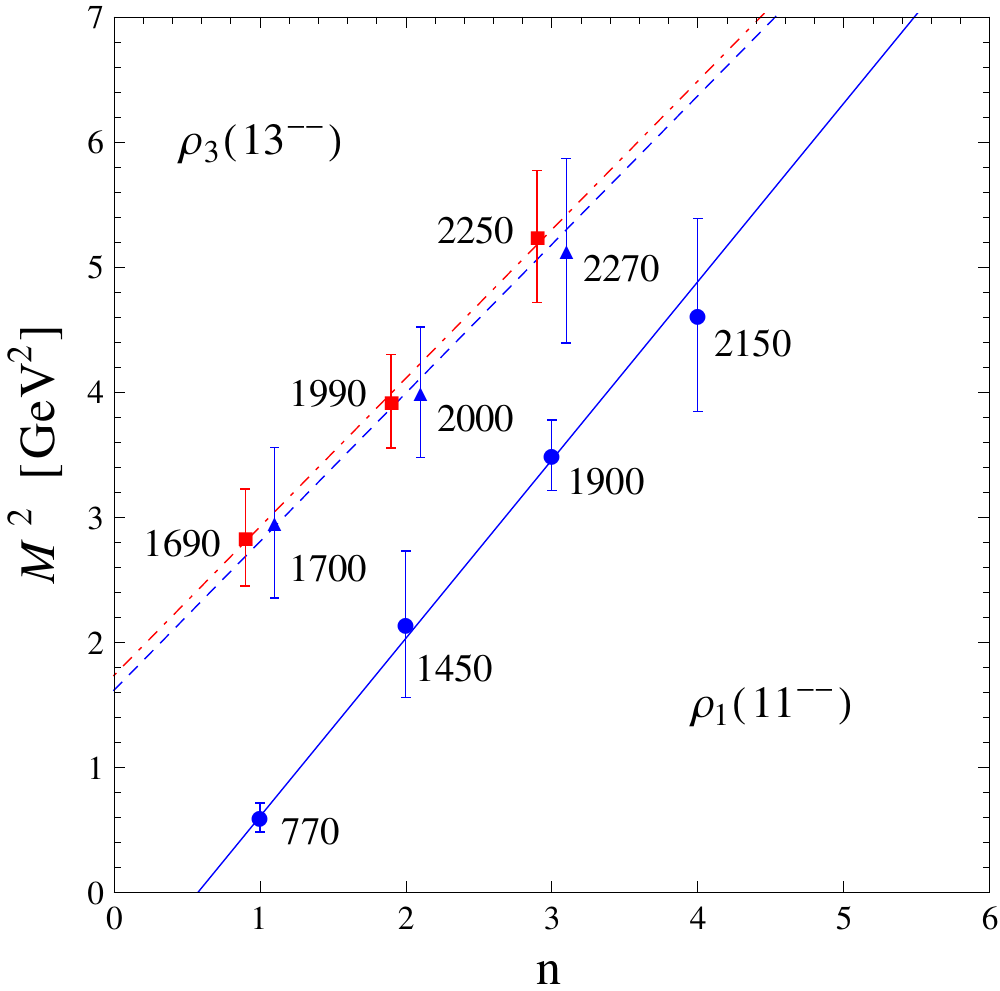}
  \caption{(color online) The $(n,M^2)$ plots for the states $\rho_1 (11^{--})$
(solid and dashed lines, respectively), and the $\rho_3 (13^{--})$ states (dot-dashed line). Error bars correspond to take $\Delta M^2=\pm \Gamma M$. \label{rhoplot}}
\end{figure}

The two trajectories for $\rho_1(11^{--})$ are depicted in
Fig.~\ref{rhoplot}.
The first one contains
$\rho(770)$, $\rho(1450)$, $\rho(1900)$, and $\rho(2150)$. As
explained by the PDG, it is not clear what values for the mass and
width one should use for $\rho(1900)$. We choose $M=1.870(30)$~GeV and
$\Gamma=0.150(20)$~GeV.  The linear fit (solid line in
Fig.~\ref{rhoplot}) yields $\mu^2=1.43(13)~{\rm GeV}^2$ with $\chi^2/{\rm DOF}=0.09$.

The second trajectory contains $\rho(1700)$, $\rho(2000)$, and
$\rho(2270)$. These last two states were predicted by
Ref.~\cite{Anisovich:2000kxa} and now are listed in the PDG
compilation. The $\rho(1700)$ and $\rho(2000)$ states, however, are
controversial and need confirmation. The corresponding slope trajectory is  $\mu^2=1.08(47)~{\rm GeV}^2$ with a $\chi^2/{\rm DOF}=0.004$ although it is drawn in Fig.~\ref{rhoplot} as parallel to the $q\bar{q}$ trajectory due to the lack of confirmation of these states. There is a new state in the
PDG Tables called $\rho(1570)$, which also needs further confirmation
because it might reflect a threshold effect or an OZI suppressed decay
mode of the $\rho(1700)$ (see the mini-review about this issue on
PDG). We do not include it on our analysis, either.

In addition to Ref.~\cite{Anisovich:2000kxa}, we have also considered the
$\rho_3(13^{--})$ states, which include $\rho_3(1690)$,
$\rho_3(1990)$, and $\rho_3(2250)$. The slope for this trajectory is
$\mu^2=1.19(32)~{\rm GeV}^2$ with $\chi^2/{\rm DOF}=0.05$. Neither $\rho_3(1690)$
nor $\rho_3(2250)$ are well established resonances and we just quote
them for completeness. We do not use the slope prediction for this
trajectory in our later average result.

\subsection{$\pi$ and $\pi_2$}

\begin{figure}
\centering
  \includegraphics[width=0.45\textwidth]{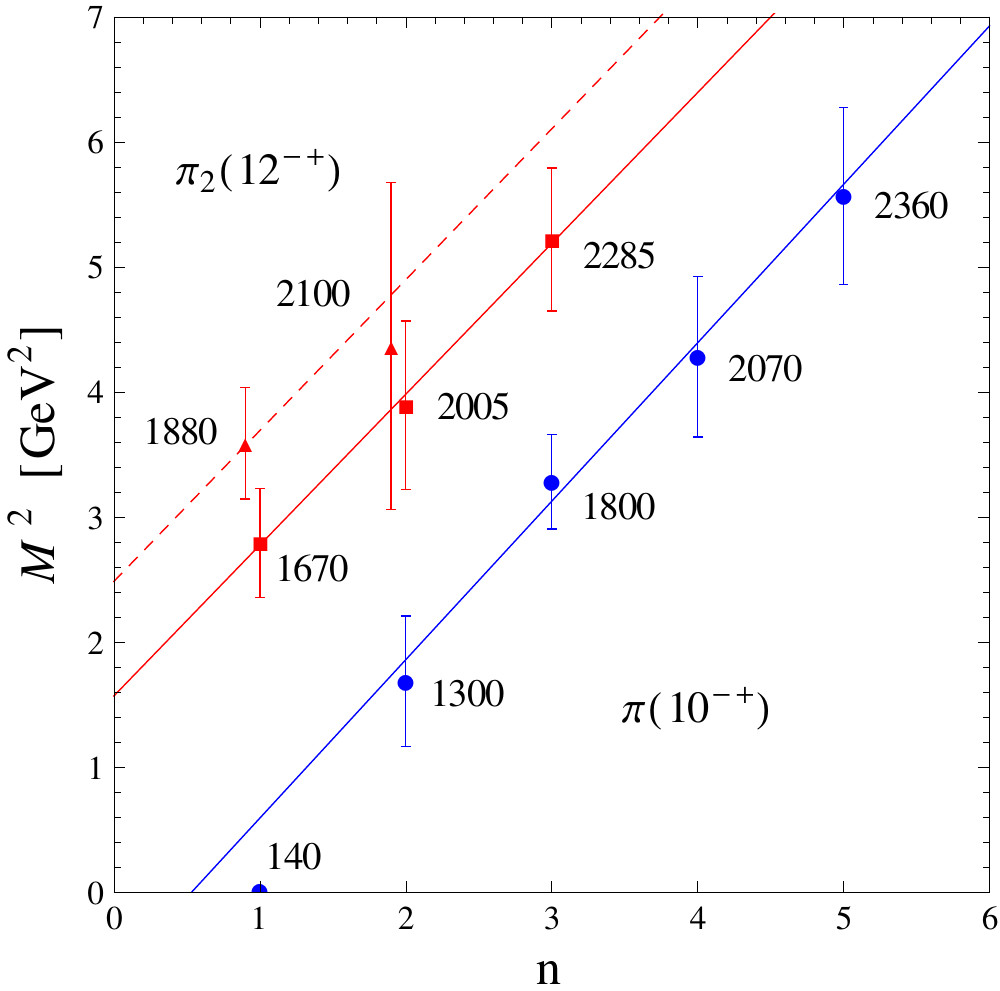}
  \caption{(color online) The $(n,M^2)$ plots for the $\pi (1 0^{-+})$
    (lower solid line) and $\pi_2 (1 2^{-+})$ (upper solid and dashed
    lines) trajectories. Error bars correspond to take $\Delta M^2=\pm \Gamma M$. \label{piplot}}
\end{figure}

For the $\pi(10^{-+})$ trajectory (lower solid line in
Fig.~\ref{piplot}), composed of $\pi(1300)$, $\pi(1800)$, $\pi(2070)$,
and $\pi(2360)$, the fit produces $\mu^2=1.27(27) {\rm GeV}^2$ with $\chi^2/{\rm
  DOF}=0.16$, where the stable $\pi(140)$ state is not used in the
fit~\footnote{One expects a strong non-linearity for the Goldstone
  bosons, see, e.g., Ref.~\cite{Arriola:2011en}.}.  We update the
trajectory including the $\pi(2070)$ and $\pi(2360)$ states originally
predicted in Ref.~\cite{Anisovich:2000kxa}.

The $\pi_2(12^{-+})$ states produce two trajectories. The first one
includes $\pi_2(1670)$ and two new states predicted in
Ref.~\cite{Anisovich:2000kxa}: $\pi_2(2005)$ and
$\pi_2(2285)$. The fit yields $\mu^2=1.21(36)~{\rm GeV}^2$ with $\chi^2/{\rm DOF}=0.02$
(solid-upper line in Fig.~\ref{piplot}). The second trajectory has
two new states, not predicted in Ref.~\cite{Anisovich:2000kxa}:
$\pi_2(1880)$ and $\pi_2(2100)$. The heaviest $\pi_2$, with the mass
of $2.090(29)$~GeV and the width of $0.625(50)$~GeV, has still to be
confirmed. Conversely, if we use the fitted daughter trajectory with
this state omitted, we predict its mass to be around $2.19(13)$~GeV.

\subsection{$a_0(10^{++})$, $a_2(12^{++})$, and $a_4(14^{++})$}

\begin{figure}
\centering
  \includegraphics[width=0.45\textwidth]{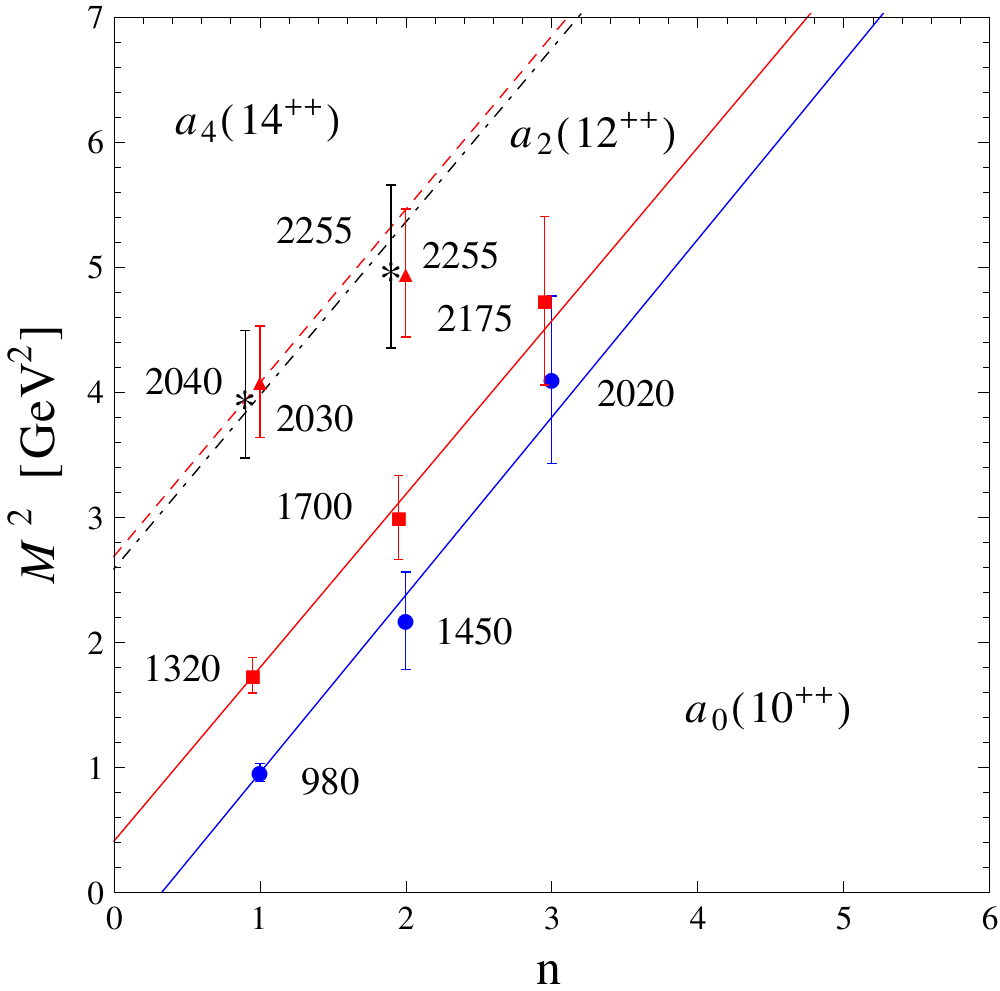}
  \caption{(color online) The $(n,M^2)$ plots for the $a_0 (10^{++})$ (lower solid line), $a_2 (12^{++})$ $q\bar{q}$ and $s\bar{s}$
(upper solid and dashed lines, respectively), and $a_4(14^{++})$ (dot-dashed line) trajectories. Error bars correspond to take $\Delta M^2=\pm \Gamma M$. \label{a024plot}}
\end{figure}

In Ref.~\cite{Anisovich:2000kxa} the experimental information in the
$a_0(10^{++})$, $a_2(12^{++})$, and $a_4(14^{++})$ sector was scarce
and could not fix the $\mu^2$ slope uniquely. Therefore, two different
slopes, $\mu^2=1.38$~GeV$^2$ and $\mu^2=1.1$~GeV$^2$ where deduced
depending on the states included, and in fact $\mu^2=1.1$~GeV$^2$
predicted a yet unobserved new state $a_0(1800)$.  Hence
$\mu^2=1.38$~GeV$^2$ is favored currently and we accept the
classification of Ref.~\cite{Anisovich:2000kxa}, with the
$a_0(10^{++})$, $a_2(12^{++})$, and $a_4(14^{++})$ trajectories and
the $a_2(12^{++})$ split into two daughters.
The $a_0(10^{++})$ contains $a_0(980)$, $a_0(1450)$\footnote{Called before $a_0(1520)$.}, and
$a_0(2020)$. The prediction for the slope is $\mu^2=1.42(26)$~GeV$^2$ with $\chi^2/{\rm DOF}=0.48$. The $a_0(2260)$ predicted in
Ref.~\cite{Anisovich:2000kxa} has not been seen yet and in our present description should be located around $2.29(12)$~GeV.

Two trajectories for $a_2(12^{++})$ are presented. The lower trajectory
contains $a_2(1320)$, $a_2 (1700)$, and $a_2(2175)$\footnote{Named
  before $a_2(1660)$ and $a_2(2100)$, respectively.}, giving
$\mu^2=1.39(26)$~GeV$^2$ with a $\chi^2/{\rm DOF}=0.24$. The
previously predicted $a_2(2400)$ Ref.~\cite{Anisovich:2000kxa} has not
been seen yet. It is also predicted within our trajectory to have
$M=2.42(17)$~GeV.

The upper trajectory contains $a_2(2030)$ and $a_2(2255)$. The
$a_2(2030)$ is an average of different experimental determinations
(under two different names) from the PDG compilation. In the 1999 PDG
edition a state called $a_2(1990)$ was introduced, while in 2001 this
state was updated to become $a_2(2030)$ by
Ref.~\cite{Anisovich:2011xe}, but not modified in the PDG review. Since then
$a_2(2030)$ appears under two different entries in the PDG
compilation, hence one of them is redundant.  The mass and width for
$a_2(2030)$ are not averaged by the PDG, where just the three
different measurements are presented. We average them with the result
$2021(14)$~MeV for the mass and $220(23)$~MeV for the width. This trajectory would produce $\mu^2=1.0(7)$~GeV$^2$.

Another problem to face for the upper $a_2$ trajectory is the presence
of two very close resonances, $a_2(1950)$ with the mass of
$1950(50)$~MeV and the width of $187(50)$~MeV, and $a_2(2030)$. It is
argued in Ref.~\cite{Anisovich:2011xe} that it is necessary to obtain
a better fit to the data. As a mater of fact, due to the large errors
of the mass position and widths, these states might easily be a single
state.  For the presented reasons, we do not use this trajectory for
our average slope value.

Finally, the $a_4(14^{++})$ trajectory contains $a_4(2040)$ and
$a_4(2255)$ and the slope turns out to be $\mu^2=1.0(8)$~GeV$^2$. We draw, however, a parallel line to the main (solid) trajectory in Fig.~\ref{a024plot}.

\subsection{$f_2(02^{++})$}

\begin{figure*}
\centering
  \includegraphics[width=0.45\textwidth]{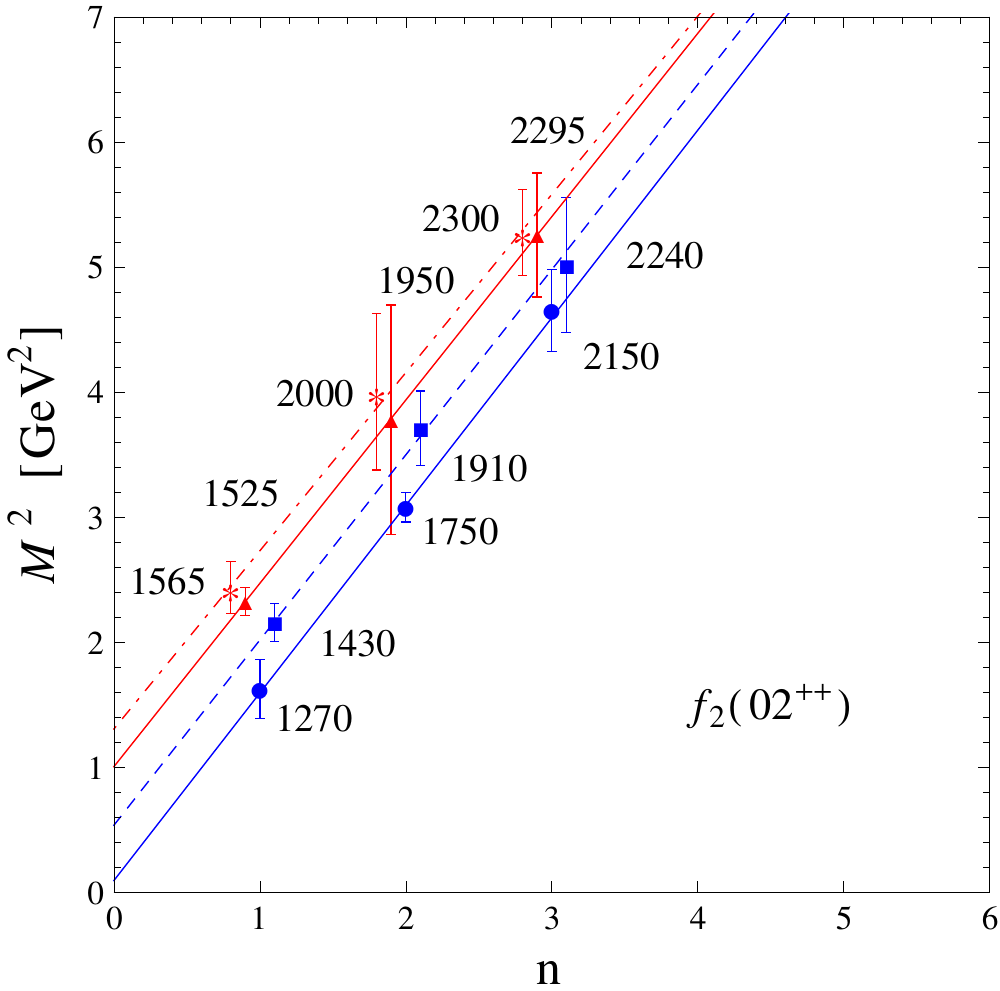}
  \includegraphics[width=0.45\textwidth]{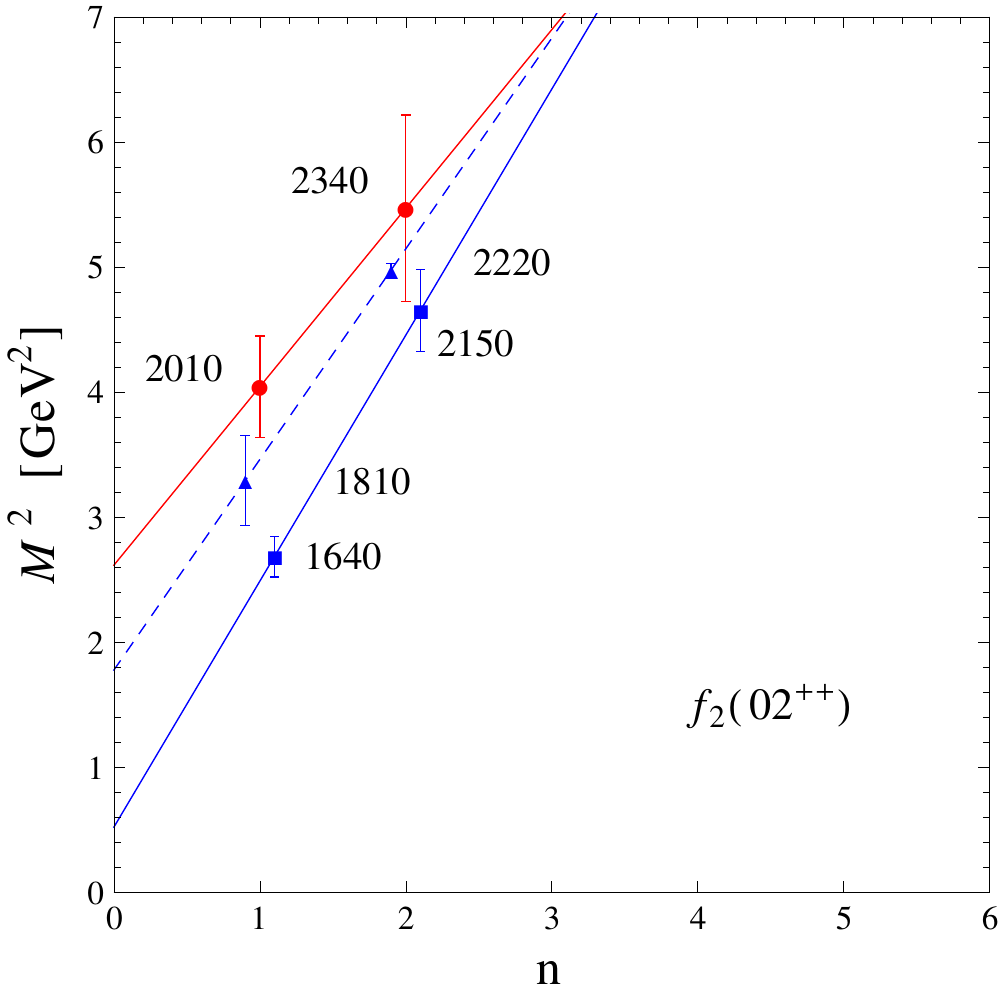}
  \caption{(color online) The $(n,M^2)$ plots for the seven $f_2
    (02^{++})$ trajectories comprising the 4 originally described in
    Ref.~\cite{Anisovich:2000kxa} (left panel) and the new ones
    discussed in the main text (right panel). Error bars correspond to take $\Delta M^2=\pm \Gamma M$.
\label{f2plot}}
\end{figure*}

In Ref.~\cite{Anisovich:2000kxa} it was not possible to discriminate
the slopes $\mu^2=1.1$~GeV$^2$ or $\mu^2=1.38$~GeV$^2$ for the
$f_2(02^{++})$ sector with the 12 available states in year 2000. The
currently listed 18 states favor the second slope.

Figure $2(c)$ of Ref. \cite{Anisovich:2000kxa} shows a quadruplet of
trajectories wtih two flavor components, $q\bar{q}$ and
$s\bar{s}$. With the inclusion of the additional 6 new states, we find
that an overall satisfactory update of Ref.~\cite{Anisovich:2000kxa}
is given by the scheme presented in Fig.~\ref{f2plot} (left panel)
requiring some reshuffling which we describe below. We name these
trajectories, $f_2^a$,$f_2^b$, $f_2^c$ and $f_2^d$.

\begin{itemize}

\item The $f_2^a$ trajectory (lower solid line in
  Fig.~\ref{f2plot}). It contains $f_2(1270)$, $f_2(1750)$, and
  $f_2(2150)$, one less state than Ref.~\cite{Anisovich:2000kxa} which
  also included $f_2(2400)$. In our case, the slope for this
  trajectory is $\mu^2=1.50(19)$~GeV$^2$ with $\chi^2/{\rm DOF}=0.06$.

\item The $f_2^b$ trajectory (dashed line in Fig.~\ref{f2plot}).
  It contains $f_2(1430)$, which is still to be determined (we take
  $M=1468(60)$~MeV and $\Gamma=100(100)$~MeV), $f_2(1910)$ with
  $M=1927(32)$~MeV and $\Gamma=154(73)$~MeV, and, finally,
  $f_2(2240)$. These states yield $\mu^2=1.48(23)$~GeV$^2$ with
  $\chi^2/{\rm DOF}=0.09$.

\item The $f_2^c$ trajectory (the upper solid line in Fig.~\ref{f2plot}.)
As in Ref.~\cite{Anisovich:2000kxa}, it
is composed of $f_2(1525)$, $f_2(1950)$, and $f_2(2295)$, giving
$\mu^2=1.47(25)$~GeV$^2$ with $\chi^2/{\rm DOF}=0.00001$.

\item The $f_2^d$ trajectory (the dot-dashed line in
  Fig.~\ref{f2plot}). It contains $f_2(1565)$ (which needs
  confirmation), $f_2(2000)$, and $f_2(2300)$. The slope for this
  trajectory is $\mu^2=1.42(20)$~GeV$^2$ with $\chi^2/{\rm DOF}=0.05$.
  The states considered here involve some reshuffling compared to
  Ref.~\cite{Anisovich:2000kxa}.

\end{itemize}

We now turn to the new trajectories, i.e.,  not given in
Ref.~\cite{Anisovich:2000kxa}, which are separately plotted in
Fig.~\ref{f2plot} (right panel).

\begin{itemize}

\item The two trajectories including $f_2(1640)$ and $f_2(2150)$ as well as $f_2(1810)$ and $f_2(2220)$ (both
  of them need confirmation), might actually be intertwined or describe an overcomplete set of
  states. The first one returns $\mu^2=1.99(36)$~GeV$^2$ and the second  $\mu^2=1.69(36)$~GeV$^2$. Considering the lack of confirmation and the particular
  values for both masses and widths, it might turn out that
  $f_2(1810)$ and $f_2(2220)$ are the very same $f_2(1910)$ and
  $f_2(2240)$ states.

\item The upper trajectory is described with $\mu^2=1.43(83)$~GeV$^2$ and contains two states, $f_2(2010)$ and
$f_2(2340)$.

\end{itemize}

\subsection{$f_0(00^{++})$}

\begin{figure}
\centering
  \includegraphics[width=0.45\textwidth]{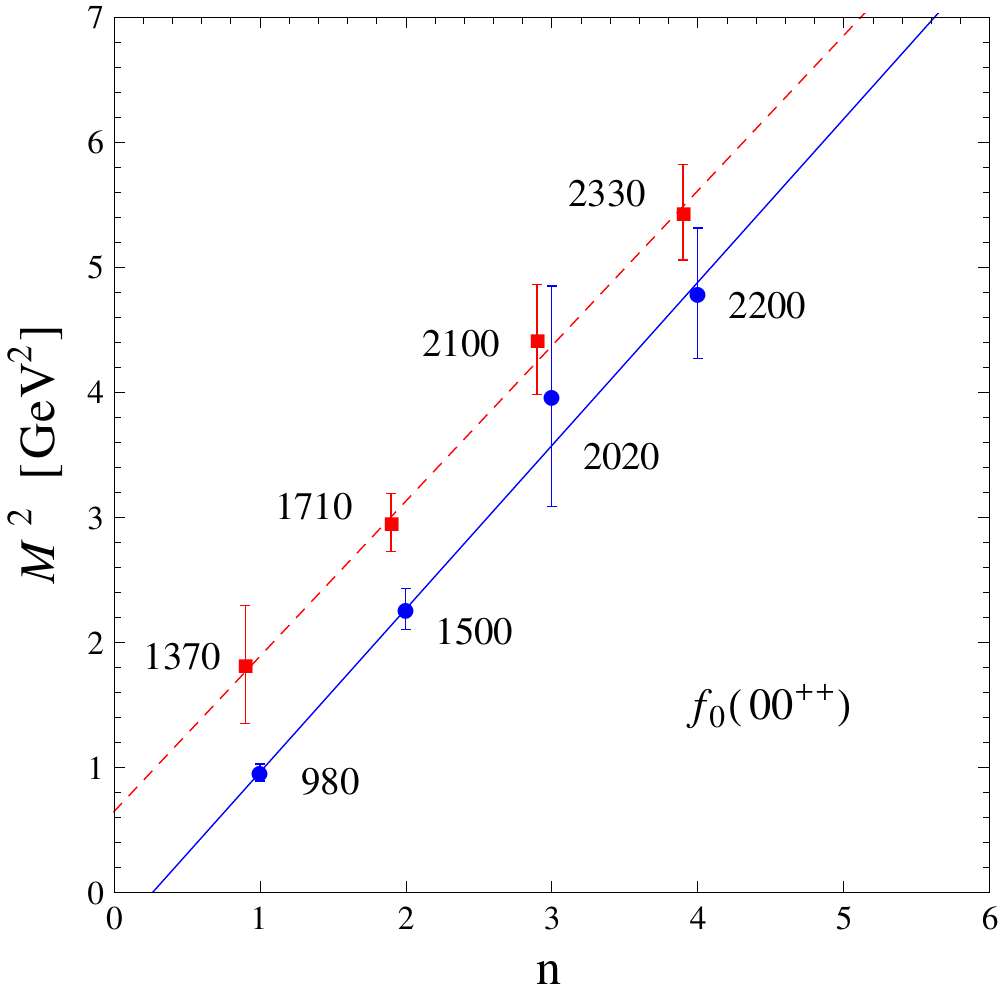}
  \caption{(color online) The $(n,M^2)$ plot for the $f_0 (00^{++})$ $q\bar{q}$ (solid line) and $s\bar{s}$ (dashed line) trajectories. Error bars correspond to take $\Delta M^2=\pm \Gamma M$. \label{f0plot}}
\end{figure}

Two trajectories for $f_0(00^{++})$ are displayed in
Fig.~\ref{f0plot} and as claimed in Ref.~\cite{Anisovich:2000kxa}
they are doubled due to two flavor components, $q\bar{q}$ and
$s\bar{s}$. Without considering the $f_0(600)$ (see however
Ref.~\cite{RuizArriola:2010fj} and Appendix \ref{sec:app}), also called the $\sigma$ meson, the
lower trajectory contains four states: $f_0(980)$, $f_0(1500)$,
$f_0(2020)$, and $f_0(2200)$ (solid line in Fig.~\ref{f0plot}). The
last state was actually predicted in Ref.~\cite{Anisovich:2000kxa} and
later confirmed experimentally. The trajectory yields
$\mu^2=1.31(12)$~GeV$^2$ with $\chi^2/{\rm DOF}=0.11$.

The second trajectory (dashed line in Fig.~\ref{f0plot}) has also
four states, $f_0(1370)$, $f_0(1710)$, $f_0(2100)$, and $f_0(2330)$,
where an average of the experimental determinations is considered for
this latter state. It yields $\mu^2=1.24(18)$~GeV$^2$ with
$\chi^2/{\rm DOF}=0.12$.

\subsection{$\omega(01^{--})$ and $\omega_3(03^{--})$}
\label{sec:radial-Regge2}

After reanalyzing the radial Regge trajectories taken into account in
Ref.~\cite{Anisovich:2000kxa}, we now analyze using the same
methodology the remaining meson families included in the latest
PDG review~\cite{Nakamura:2010zzi}.

\begin{figure}
\centering
  \includegraphics[width=0.45\textwidth]{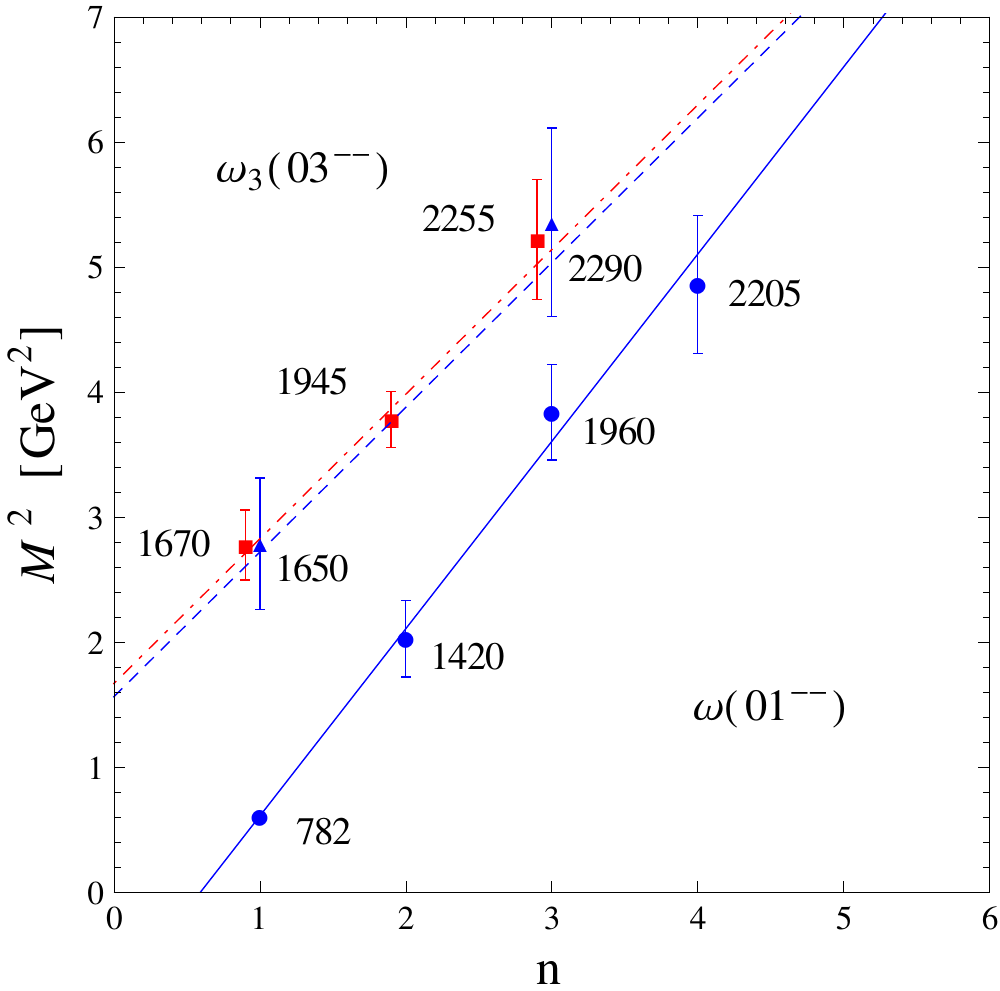}
  \caption{(color online) The $(n,M^2)$ plot for the $\omega(01^{--})$  (circles, solid line and triangles, dashed line) trajectories,
             as well as for the $\omega_3(03^{--})$ (squares, dot-dashed line) trajectory. Error bars correspond to take $\Delta M^2=\pm \Gamma M$. \label{wplot}}
\end{figure}

Two trajectories for the $\omega(01^{--})$ states are shown in
Fig.~\ref{wplot}.  The ordering of the states on
the different trajectories for the $\omega$-family follows very
closely the classification of the $\rho$-family states,
Fig.~\ref{rhoplot}.  The main $\omega$-trajectory, representing the
$q\bar{q}$ states and drawn as a solid line in Fig.~\ref{wplot},
contains four states: $\omega(782)$, $\omega(1420)$, $\omega(1960)$,
and $\omega(2205)$. The slope from the fit is $\mu^2=1.50(12)$~GeV$^2$
with $\chi^2/{\rm DOF}=0.32$.

The daughter trajectory (dashed line in Fig.~\ref{wplot})
contains two states: $\omega(1650)$ and $\omega(2290)$. In the PDG
recollection, two $\omega$ states with very similar masses are listed
in the $2.3$~GeV region: $\omega(2290)$ and $\omega(2330)$.  Looking
at the error determination of the parameters of these states it is not
clear to us that the two states are indeed different. We gather both
experimental results in a single entry $\omega(2290)$, which has mass
$2315(45)$~MeV and width $325(185)$~MeV. Comparing this trajectory
with the corresponding one from the $\rho$-family, we notice a missing
$\omega$ state with the mass near $2000$~MeV, indicating the state $\omega(2290)$ to be the third on its trajectory. With only two states,
the corresponding slope is  $\mu^2=1.27(47)$~GeV$^2$.

The third trajectory in Fig.~\ref{wplot}, describing the $\omega_3(03^{--})$ states,
contains $\omega_3(1670)$, $\omega_3(1945)$, and $\omega_3(2255)$. It
yields $\mu^2= 1.16(26)$~GeV$^2$ with $\chi^2/{\rm DOF} =0.37$.

\subsection{$h_1(01^{+-})$ and $h_3(03^{+-})$}

\begin{figure}
\centering
  \includegraphics[width=0.45\textwidth]{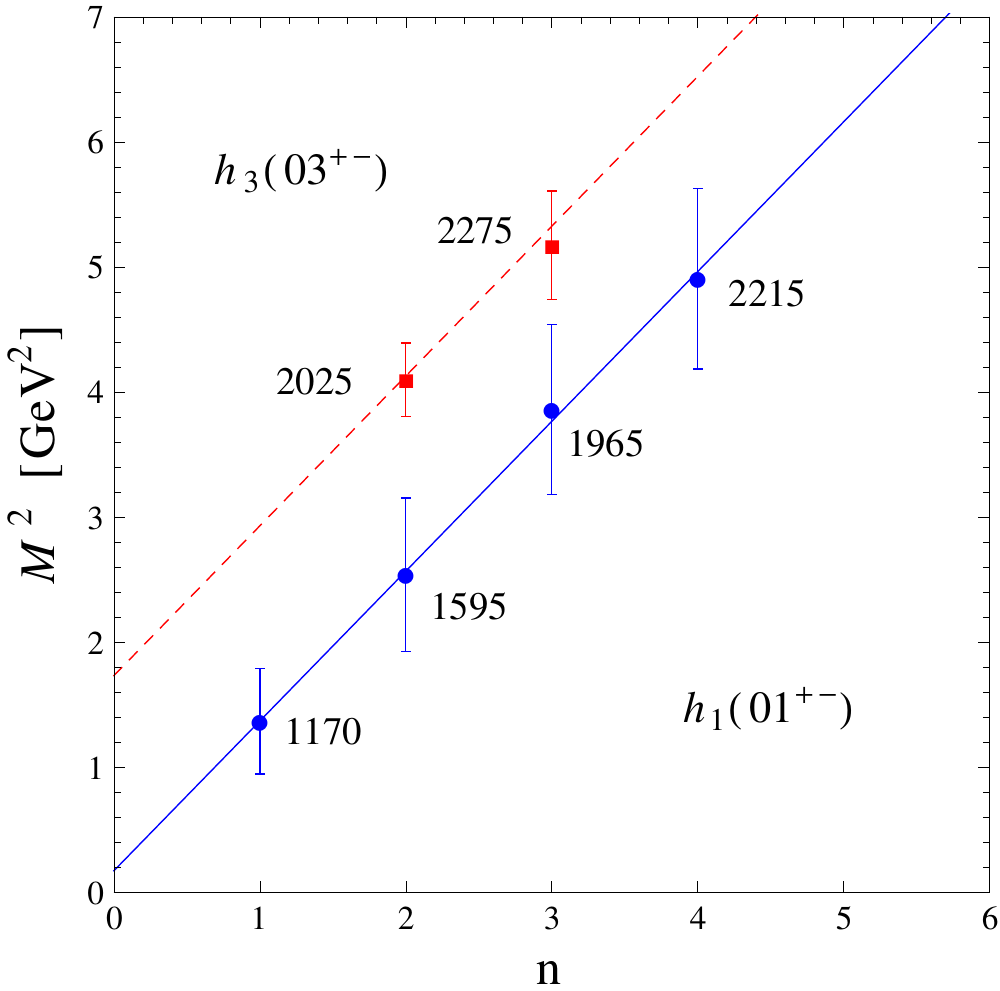}
  \caption{(color online) The $(n,M^2)$ plot for the $h_1(01^{+-})$ (circles, solid line) and $h_3(03^{+-})$ (squares, dashed line) trajectories. Error bars correspond to take $\Delta M^2=\pm \Gamma M$. \label{hplot}}
\end{figure}

The $h$ sector contains two trajectories corresponding to the
$h_1(01^{+-})$ and $h_3(03^{+-})$ states shown as circles and squares
in Fig.~\ref{hplot}, respectively.  The $h_1(01^{+-})$ case consists of
four states: $h_1(1170)$, $h_1(1595)$, $h_1(1965)$, and
$h_1(2215)$. The $h_1(1380)$ state is excluded, since it still needs to be
confirmed. The linear fit to this trajectory, shown as a solid line in
Fig.~\ref{hplot}, gives
$\mu^2=1.20(25)$~GeV$^2$ with $\chi^2/{\rm DOF}=0.01$.

The $h_3(03^{+-})$ includes only two states, $h_3(2025)$ and
$h_3(2275)$ (supposed to be the second and third excitation states of
that trajectory), thus the slope $\mu^2=1.08(54)$~GeV$^2$ is determined. In Fig.~\ref{hplot}, the
dashed line, drawn parallel to the solid line, represents this
trajectory.

\subsection{$b_1(01^{+-})$ and $b_3(03^{+-})$}

\begin{figure}
\centering
  \includegraphics[width=0.45\textwidth]{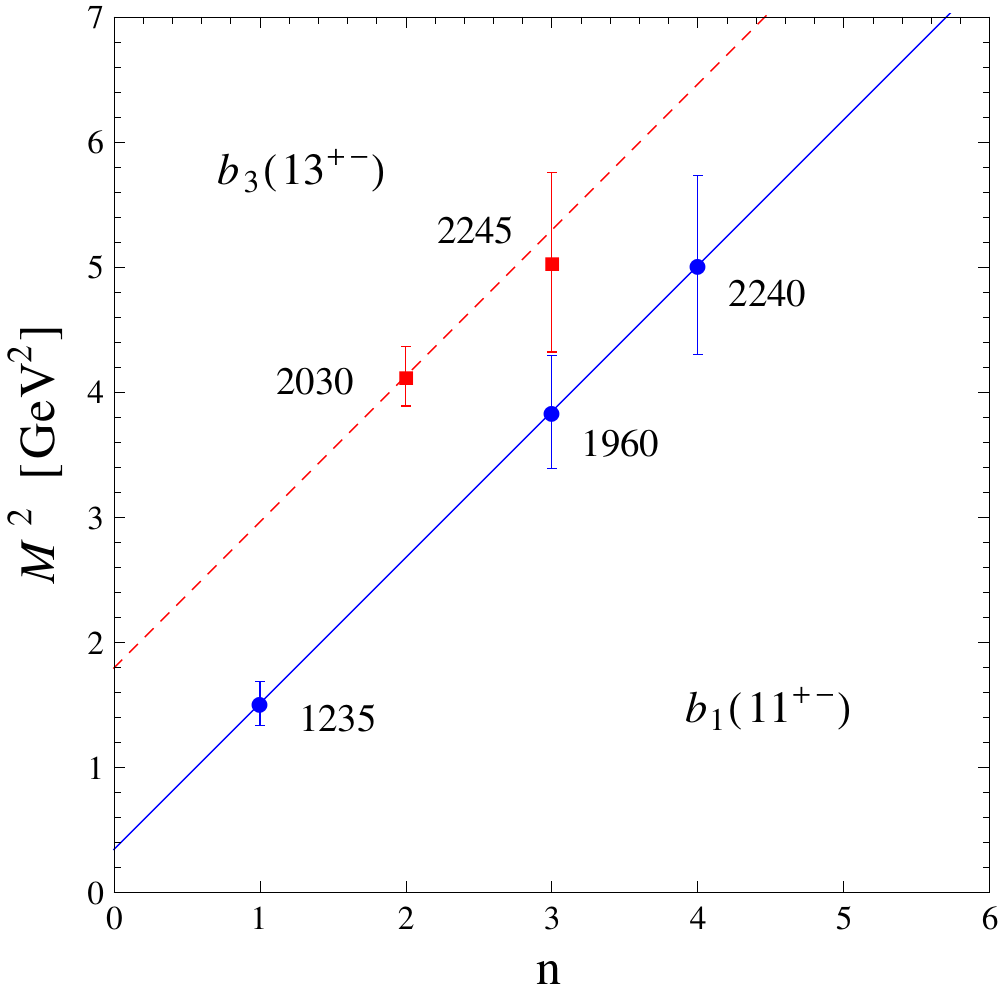}
  \caption{(color online) The $(n,M^2)$ plot for the $b_1(11^{+-})$ (circles, solid line) and $b_3(13^{+-})$ (squares, dashed line) trajectories. Error bars correspond to take $\Delta M^2=\pm \Gamma M$. \label{bplot}}
\end{figure}

Similarly to the previous subsection, the $b$ sector contains two
trajectories, corresponding to the $b_1(11^{+-})$ and $b_3(13^{+-})$
states, shown as circles and squares in Fig.~\ref{bplot},
respectively.  The $b_1(01^{+-})$ consists of three states:
$b_1(1235)$, $b_1(1960)$, and $b_1(2240)$. The linear fit returns
$\mu^2=1.17(18)$~GeV$^2$ with $\chi^2/{\rm DOF}=0.00001$ (solid line in Fig.~\ref{bplot}).

The $b_3(03^{+-})$ includes only two states, $b_3(2030)$ and
$b_3(2245)$, hence the slope is $\mu^2=0.93(75)$~GeV$^2$. In Fig.~\ref{bplot}, a
dashed line, parallel to the solid line, represents this trajectory.
Since the resemblance between the $h$ sector and the $b$ sector is
apparent, that suggests the existence of a still not determined $b_1$
state with a mass of the order of $1600$~MeV.

\subsection{$f_1(01^{++})$ and $f_3(03^{++})$}

\begin{figure}
\centering
  \includegraphics[width=0.45\textwidth]{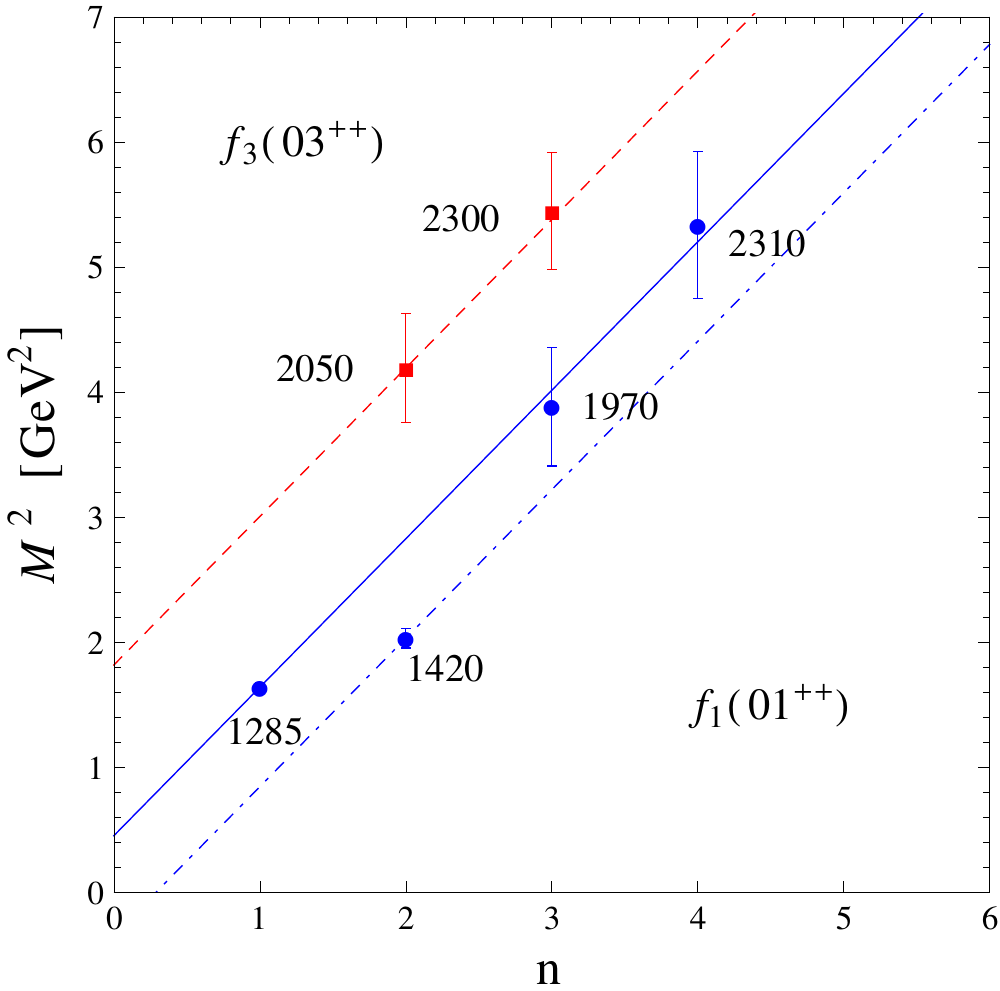}
  \caption{(color online) The $(n,M^2)$ plot for the $f_1(01^{++})$ (circles, solid line) and $f_3(03^{++})$ (squares, dashed line) trajectories. Error bars correspond to take $\Delta M^2=\pm \Gamma M$. \label{f13plot}}
\end{figure}

The situation with the $f_1$ and $f_3$ states is equivalent to the $b$
and $h$ case, we thus have two different trajectories corresponding to
the different angular-momentum, $f_1(01^{++})$ and $f_3(03^{++})$,
shown as circles and squares in Fig.~\ref{f13plot}, respectively.  The
$f_1(01^{++})$ trajectory consists of three states: $f_1(1285)$,
$f_1(1960)$, and $f_1(2240)$. The fit for this trajectory returns
$\mu^2=1.19(15)$~GeV$^2$ with $\chi^2/{\rm DOF}=0.13$ and it is shown as a solid line in
Fig.~\ref{f13plot}.

The $f_3(03^{++})$ includes only two states, $f_3(2050)$ and
$f_3(2300)$. The slope is $\mu^2=1.27(64)$~GeV$^2$. In Fig.~\ref{f13plot},
the dashed line, drawn parallel to the solid line, displays this
trajectory. The location of these states follows closely the case of
$h_3(03^{+-})$ and $b_3(13^{+-})$, hence it starts at the radial
quantum number $n=2$. From Fig.~\ref{f13plot} it is not obvious
how to allocate the $f_1 (1420)$ state, since the departure from the
expected value seems much larger than expected from the half-width
rule, therefore we exclude it from the fit. This choice resembles the
$h_1$ case.

\subsection{$\phi(01^{--})$ and $\phi_3(03^{--})$}

\begin{figure}
\centering
  \includegraphics[width=0.45\textwidth]{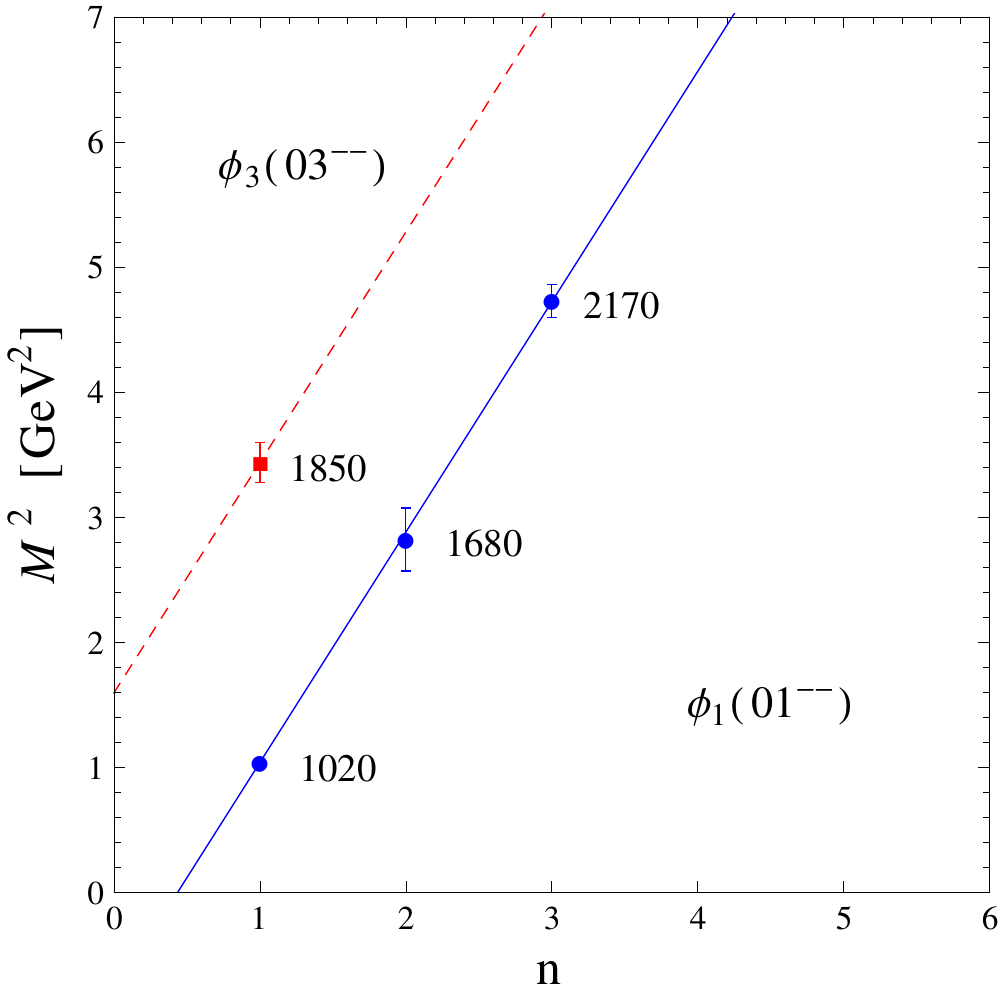}
  \caption{(color online) The $(n,M^2)$ plot for the $\phi(01^{--})$ (circles, solid line) and $\phi_3(03^{--})$ (squares, dashed line) trajectories. Error bars correspond to take $\Delta M^2=\pm \Gamma M$. \label{phi13plot}}
\end{figure}

The $\phi$ sector has three states with $J=1$
($\phi(1020)$, $\phi(1680)$, and $\phi(2170)$), and one with $J=3$
($\phi_3(1850)$), therefore the allocation of states becomes less
unique. However, if the states with $J=1$ are placed on a radial
linear trajectory, a well determined slope $\mu^2=1.84(6)$~GeV$^2$ with
a $\chi^2=0.06$ is obtained. We note that such slope is much larger
than any of the other slopes found so far. We could also concede the states $\phi(1680)$, and $\phi(2170)$ to be $n=2,3$ respectively, which would produce $\mu^2=1.19(4)$~GeV$^2$ instead, although with too large $\chi^2/{\rm DOF}=6.4$. Due to this ambiguity, we
will not consider this family for the final summary results

In Fig.~\ref{phi13plot}, two trajectories are shown as parallel lines,
with a solid line representing the $J=1$ states, and the dashed line going across the
the $J=3$ state. Clearly, this somewhat disturbing
picture should profit from both theoretical or experimental insight.

\subsection{$\pi_1(11^{-+})$}

Finally, the last sector we analyze corresponds to the $\pi_1$ states,
composed by $\pi_1(1400)$, $\pi_1(1600)$, and $\pi_1(2015)$.  Two
different measurements are found for this last state and we average
them to have mass $M=2.013(25)$~GeV and width $\Gamma=0.287(53)$~GeV.

\begin{figure}
\centering
  \includegraphics[width=0.45\textwidth]{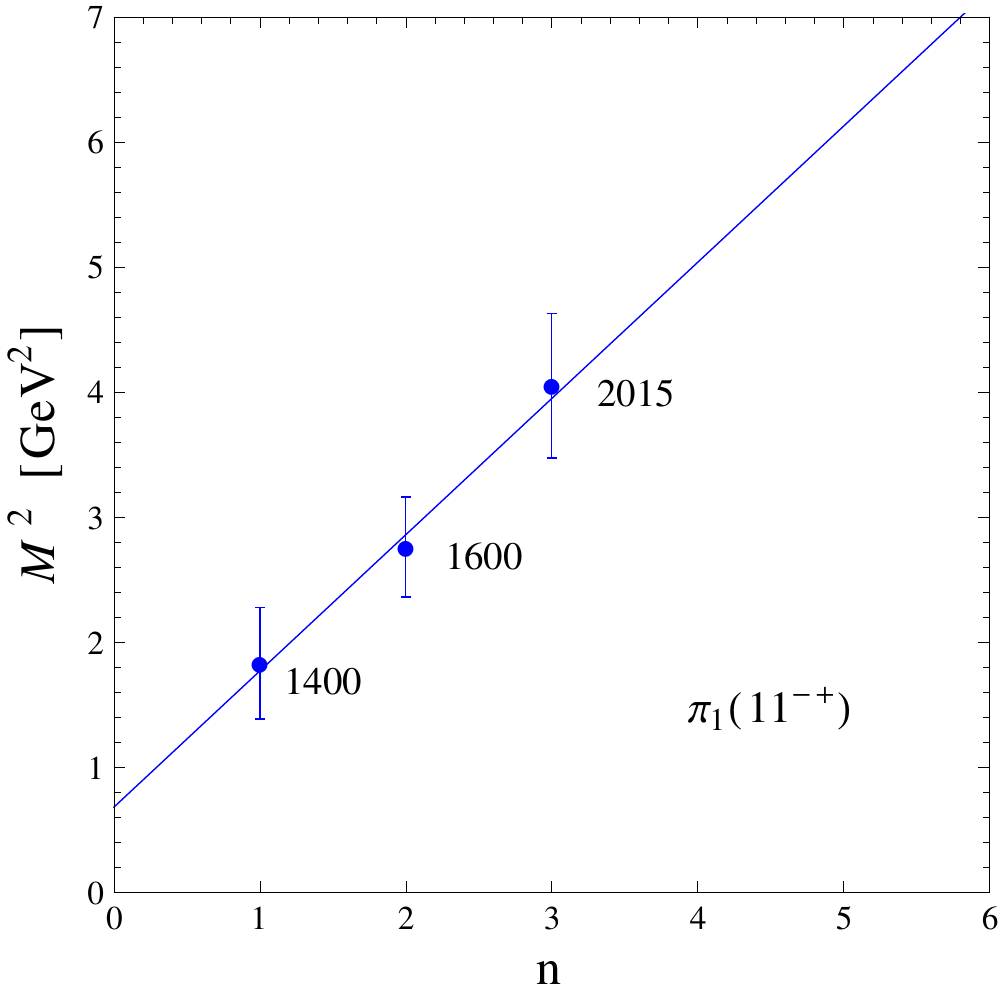}
  \caption{(color online) The $(n,M^2)$ plot for the $\pi_1(11^{-+})$
(circles, solid line)  trajectories. Error bars correspond to take $\Delta M^2=\pm \Gamma M$. \label{pi1plot}}
\end{figure}

With these three states on a linear trajectory the slope obtained is
$\mu^2=1.09(36)$~GeV$^2$ with a $\chi^2=0.11$. The results are
depicted in Fig.~\ref{pi1plot}.

\subsection{Summary of the radial-trajectory fits \label{sec:summary-radial}}

We summarize this Section by collecting the fits for all the radial trajectories
studied. The $\mu^2$ parameter ranges from $1.09(36)$~GeV$^2$ (the $\pi_1(11^{-+})$ trajectory) to $1.50(19)$~GeV$^2$
(corresponding to the $f_2^a(02^{++})$ trajectory).  The weighted average
yields~\footnote{We use the customary definition for the weighted
  average $\bar A = \sum_{i=1}^N w_i A_i/ \sum_{i=1}^N w_i $, with
  $w_i = 1/\sigma_i^2$. The errors are the mean-squared deviation, $\sqrt{\bar
    A^2- (\bar A)^2}$. The mean average corresponds to $w_i = 1$.}
\begin{eqnarray}
\mu^2=1.35(4){\rm ~GeV}^2. \label{eq:mu_g}
\end{eqnarray}
This result agrees within the uncertainties with the estimate of
Ref.~\cite{Anisovich:2000kxa}, $\mu^2=1.25(15)$~GeV$^2$, where the
uncertainty was given by the spread of the mean values determined from a
fit to the PDG masses with equal weights. When we carry out
the same procedure for the updated and {\it new} trajectories, we
get $\mu^2=1.32(12)$~GeV$^2$. This seems to provide a quite
robust estimate of a {\it common} radial Regge trajectories slopes.
A graphic overview of the estimated slopes is presented in Fig.~\ref{muJall} (upper part).

We have also considered the possibility of a linear n-dependence of the masses, 
since it was a popular outcome of holographic models in the hard-wall scheme 
(see e.g. Ref.~\cite{Erdmenger:2007cm} and rererences therein).
With the same conditions as analyzed above, i.e., assuming the validity of the 
half-width rule our analysis is not compatible with such radial spectrum; typically 
we obtain $\chi^2/{\rm DOF} \sim 10$ or larger. 

\section{$(J,M^2)$ trajectories \label{sec:J-Regge}}

In this section, taking into account all the states so far considered
and adding those with larger $J=4,5,6$ from the PDG
Tables~\cite{Nakamura:2010zzi}, we complement the results of the
$(n,M^2)$ analysis with the study in the $(J,M^2)$ plane, i.e, the
standard Chew-Frautschi plots~\cite{Chew:1961ev}.
One may parametrize the trajectories as
\begin{equation}\label{AnisovichAlpha}
\alpha_X(M^2)\sim \alpha_X(0)+\alpha'_X(0)M^2\, ,
\end{equation}
with $\alpha_X(0)$ and $\alpha'_X(0)$ constant parameters.
Equivalently, we consider
\begin{equation}
M_X^2(J)=M_X^2 (0)+ \beta_X J\, .
\end{equation}
The $\beta$ parameter is related to $\alpha$ as $\beta\sim 1/\alpha'_X(0)$.

\begin{figure*}
\centering
  \includegraphics[width=0.45\textwidth]{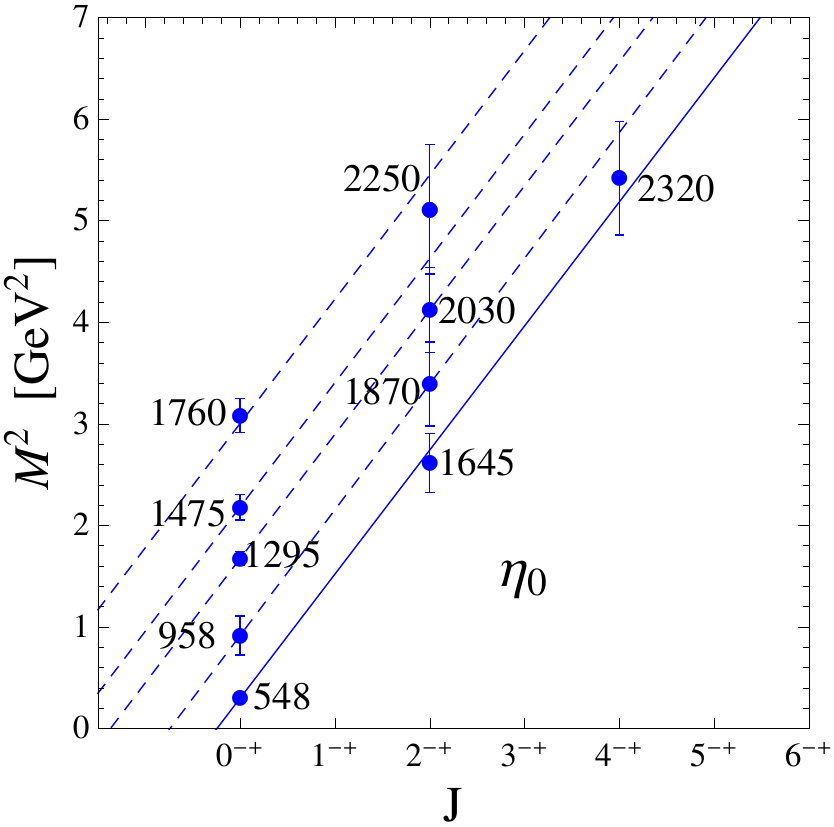}
  \includegraphics[width=0.45\textwidth]{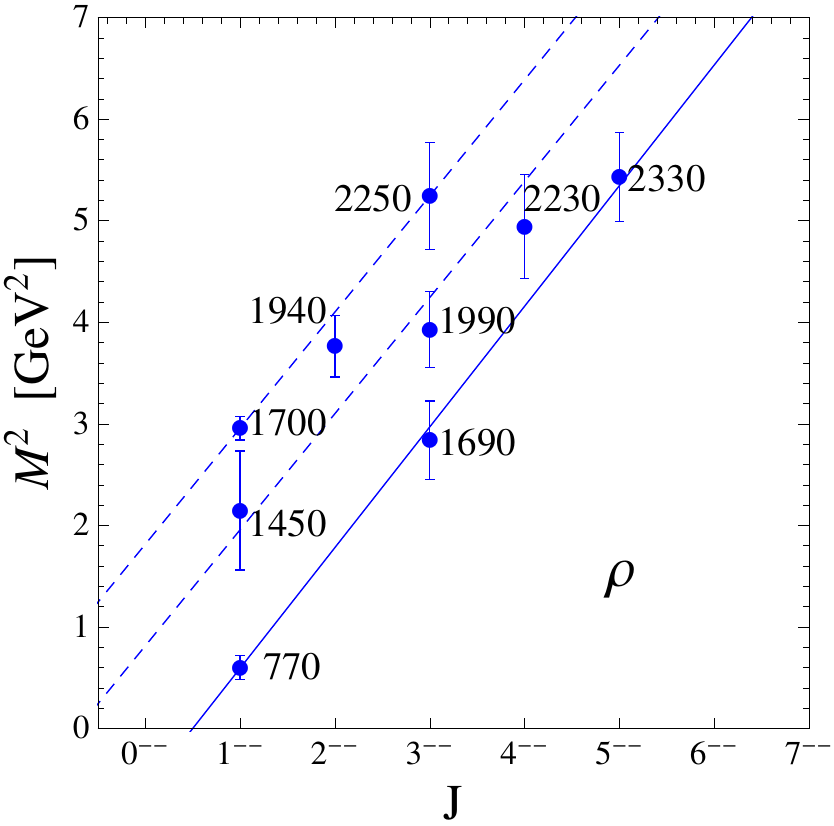}\\
  \includegraphics[width=0.45\textwidth]{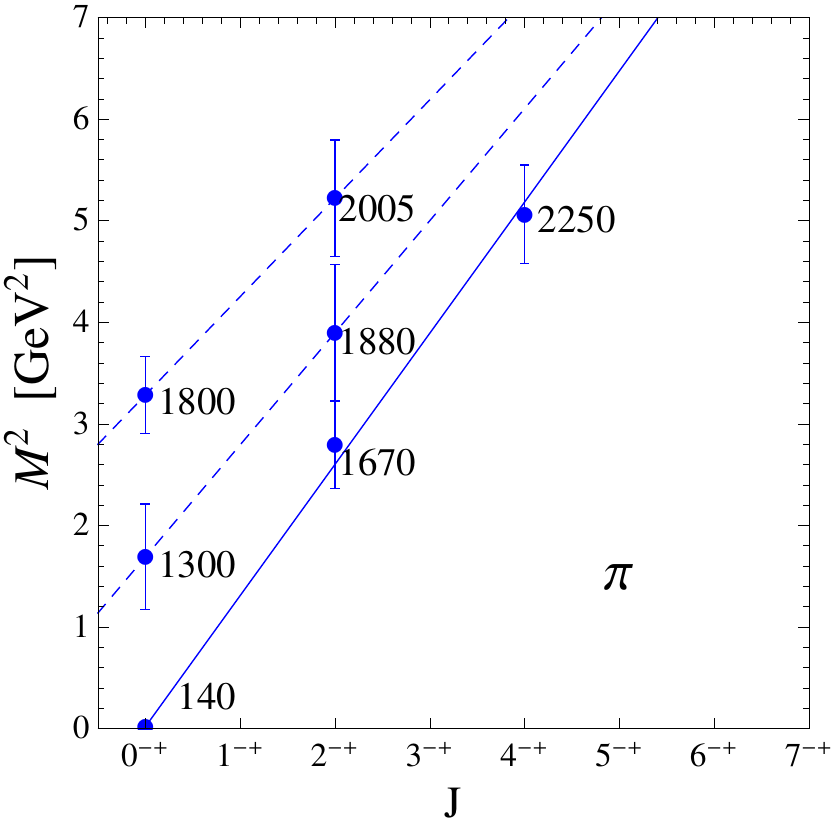}
  \includegraphics[width=0.45\textwidth]{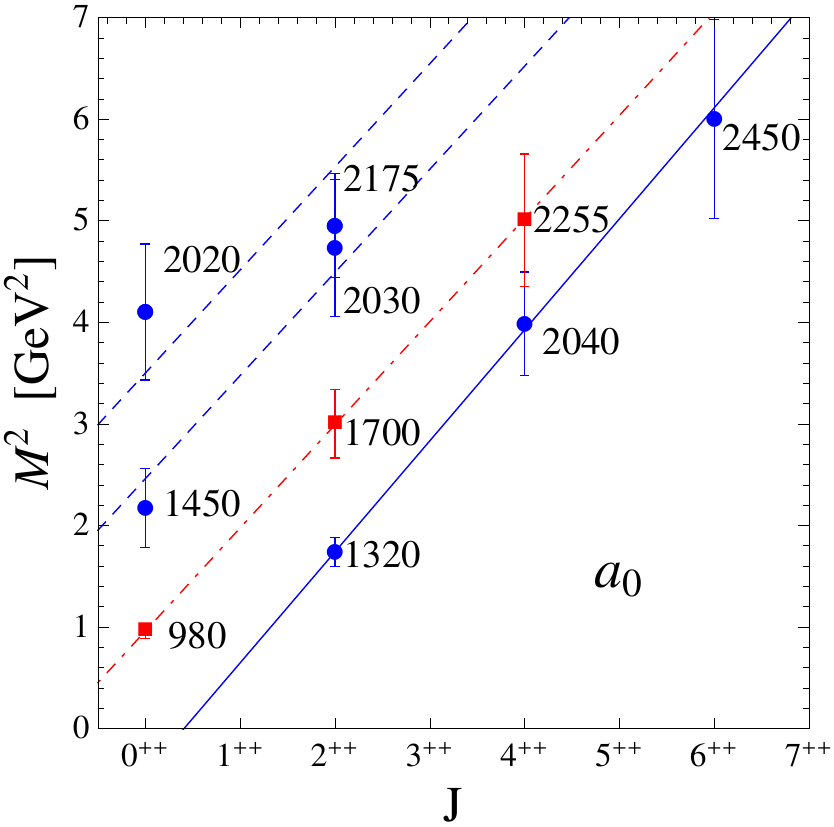}\\
  \includegraphics[width=0.45\textwidth]{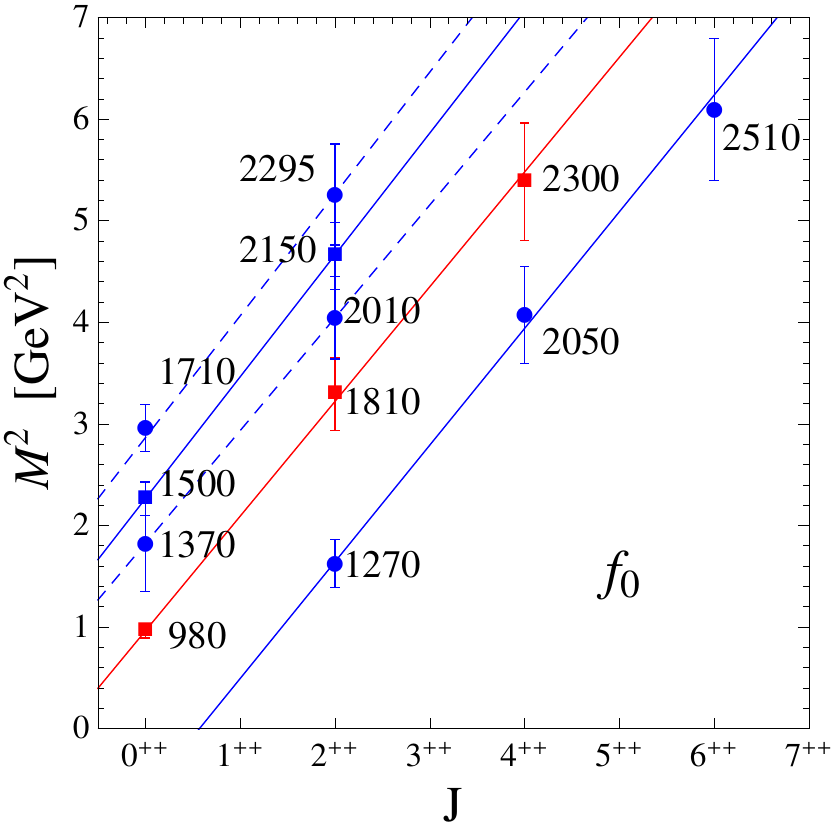}
  \includegraphics[width=0.45\textwidth]{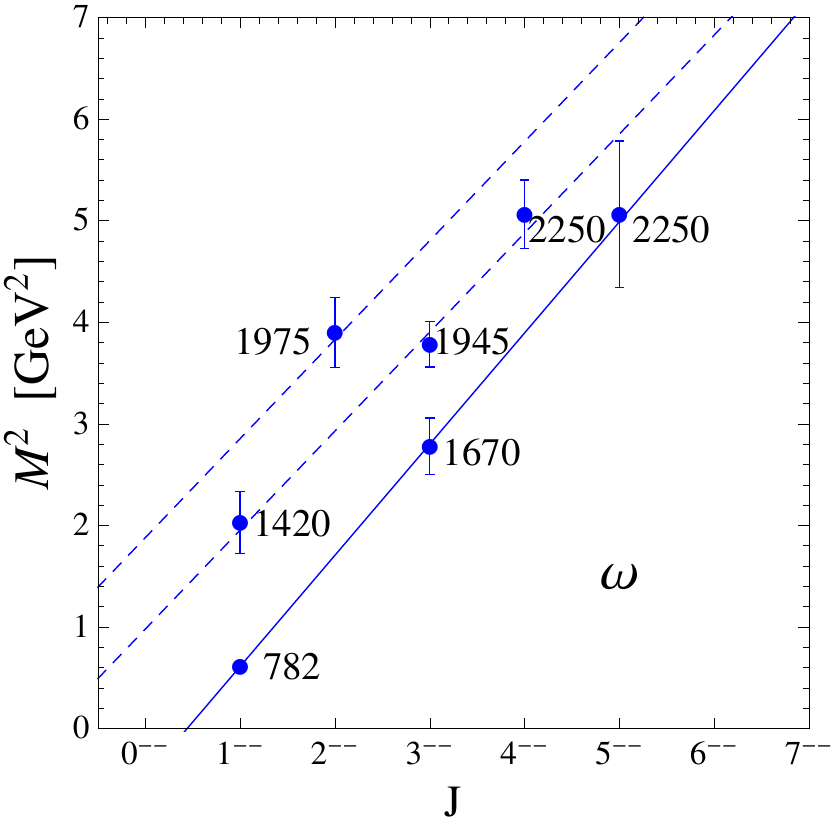}
  \caption{(color online) The $(J,M^2)$ plots for all states considered in Table~\ref{TableJ}. Error bars correspond to take $\Delta M^2=\pm \Gamma M$.\label{Jplot}}
\end{figure*}

\begin{figure}
\centering
  \includegraphics[width=0.45\textwidth]{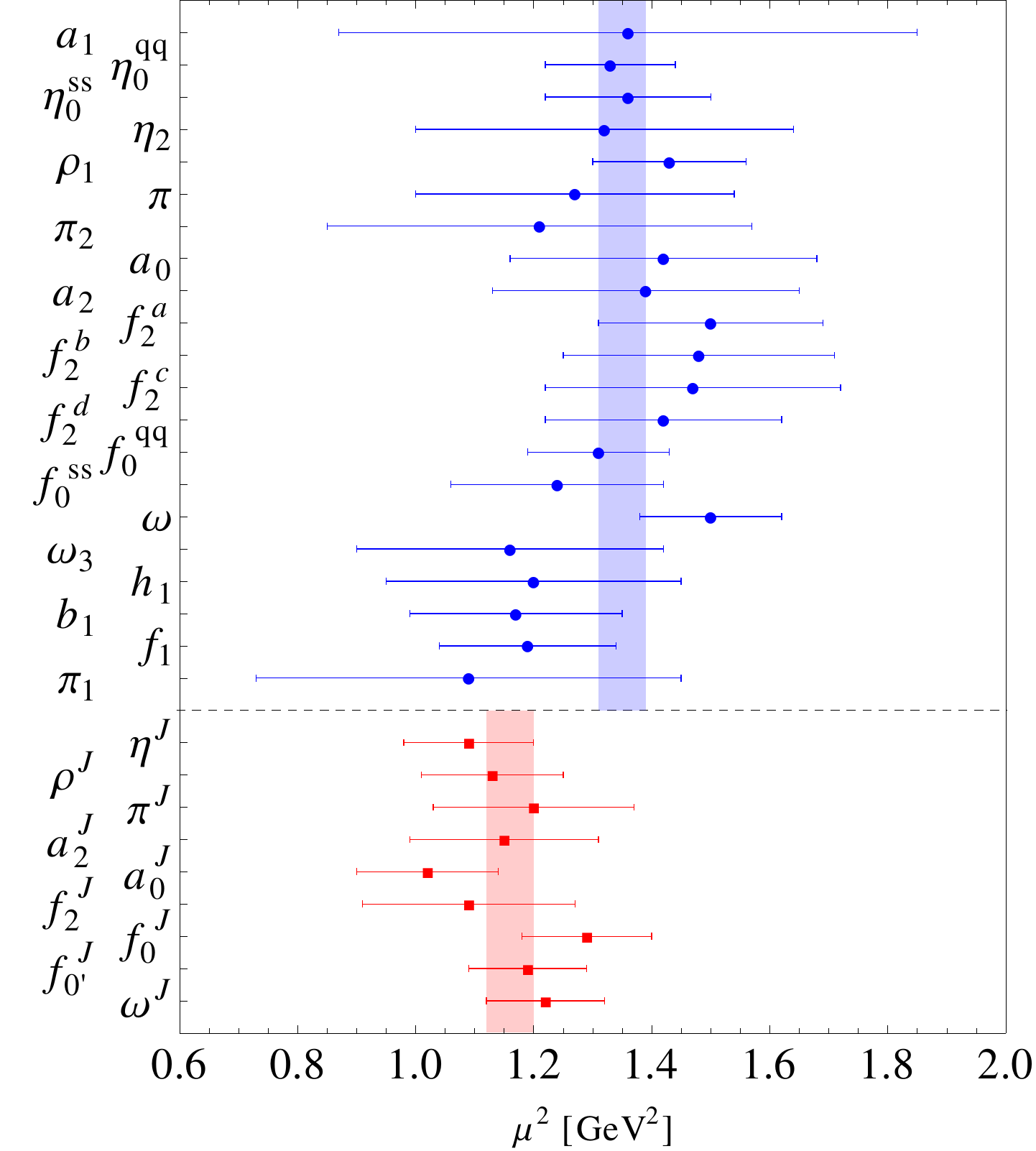}
  \caption{(color online) $(n,M^2)$ and $(J,M^2)$ slope results for
    the considered trajectories. The horizontal dashed line separates
    the radial slopes (circles) from the angular slopes
    (squares). Individual errors are estimated from the $\chi^2$ fits to the
    corresponding trajectories described in the main
    text. The bands correspond to the weighted averages of the radial (upper band)
and the angular-momentum (lower band). 
\label{muJall}}
\end{figure}

The results for the leading trajectories are shown in Table~\ref{TableJ}. Generally, we do not attempt to determine the slope when
the trajectory is made of less than three states.  An overview of the
estimated slopes can be seen in Fig.~\ref{muJall} (lower part).
Nevertheless, to provide a broader perspective, we show in
Fig.~\ref{Jplot} also trajectories with just two states (dashed
lines).

\begin{table}
  \centering
\renewcommand{\arraystretch}{1.5}
{
\begin{tabular}{|c||c|c|c|c||}
\hline
   $X$  & $M_X^2 (0)$~[GeV$^2$]  &$\beta_X$~[GeV$^2$] & $\alpha'_X(0)$~[GeV$^{-2}$]  & $\chi^2/{\rm DOF}$ \\
\hline
$\eta$  & 0.30(0) & 1.22(10)  & 0.82(7) &  0.37 \\
$\rho$  & 0.60(11) & 1.19(10) & 0.84(7) &  0.15 \\
$\pi$   & 0.018(0) & 1.29(11) & 0.78(7) &  0.26\\
$a_2$   & -0.45(43) & 1.09(18)  & 0.92(15)&  0.03 \\
$a_0$   & 0.96(7)  & 1.02(12) & 0.98(12)&  0.001\\
$f_2$   & -0.66(48)& 1.15(16)  & 0.87(12) &  0.12 \\
$f_0$   & 2.26(16)  & 1.20(17)  & 0.83(12)&  0.01 \\
$f_0'$   & 0.96(7)  & 1.13(12)  & 0.88(9)&  0.07 \\
$\omega$& -0.48(11) &  1.09(11) & 0.92(9) &  0.02 \\
\hline
\end{tabular}
}
\caption{The $(J,M^2)$ trajectories for leading and daughter trajectories. For an easy of comparison with Ref.~\cite{Anisovich:2000kxa}, we also show the corresponding $\alpha'_X(0)\sim 1/\beta$ for each trajectory.  \label{TableJ} }
\end{table}

The weighted average result for the angular trajectories yields
\begin{eqnarray}
\beta=1.16(4)~{\rm GeV}^2.
\label{eq:beta_g}
\end{eqnarray}
If one considers, instead, the spread of central values as in
Ref.~\cite{Anisovich:2000kxa}, the updated result for the states, one obtains
$\beta=1.15(8) {\rm~GeV}^2 $, in agreement with~\cite{Anisovich:2000kxa}.

When comparing the results of Eq.~(\ref{eq:mu_g}) and Eq.~(\ref{eq:beta_g}),
we note that the radial and angular-momentum slopes are different at
the level of 3.4 standard deviations. Thus, there is indication that the
{\em radial slopes are larger than the angular-momentum slopes} at a
significant statistical level. We come back to this important issue in the next section.

It should be noted that in addition to the states of
Ref.~\cite{Anisovich:2000kxa}, we have also included the $\omega$,
$h_1$, $b_1$, $f_1$, $\phi$, and $\pi_1$ sectors on our analysis, as
well as three new $J$-trajectories, with the $a_0$, $f_0$, and
$\omega$ states.

\begin{figure*}
\centering
  \includegraphics[width=0.45\textwidth]{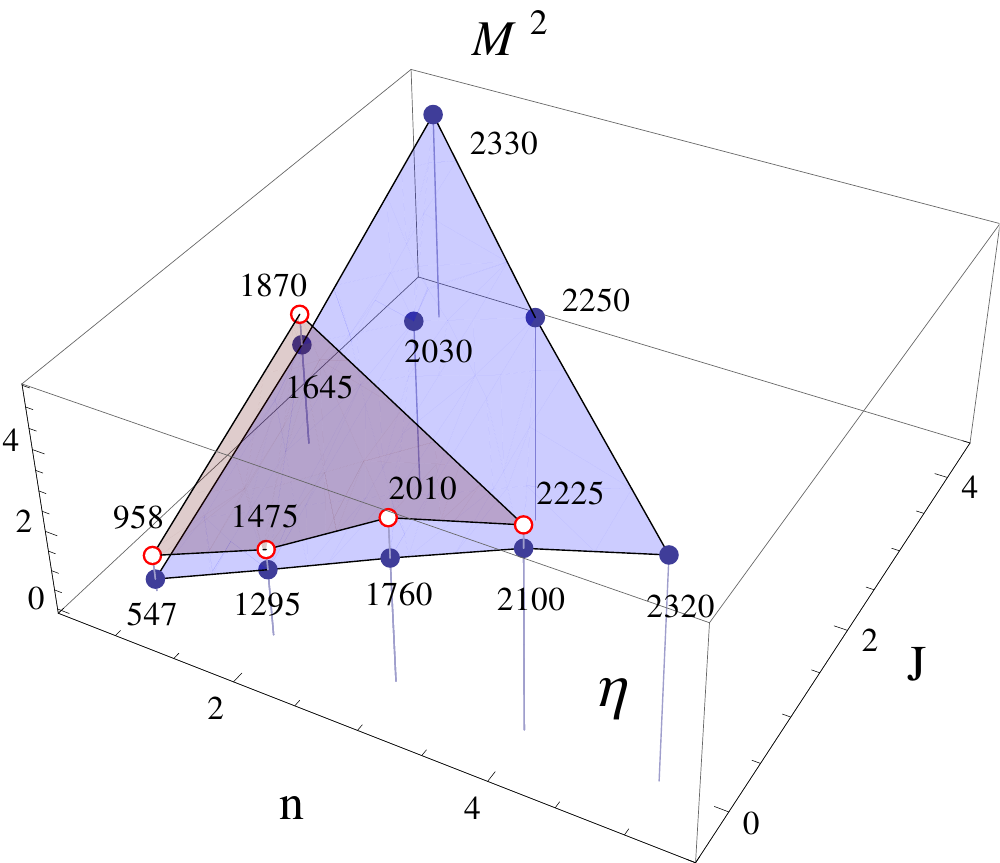}
  \includegraphics[width=0.45\textwidth]{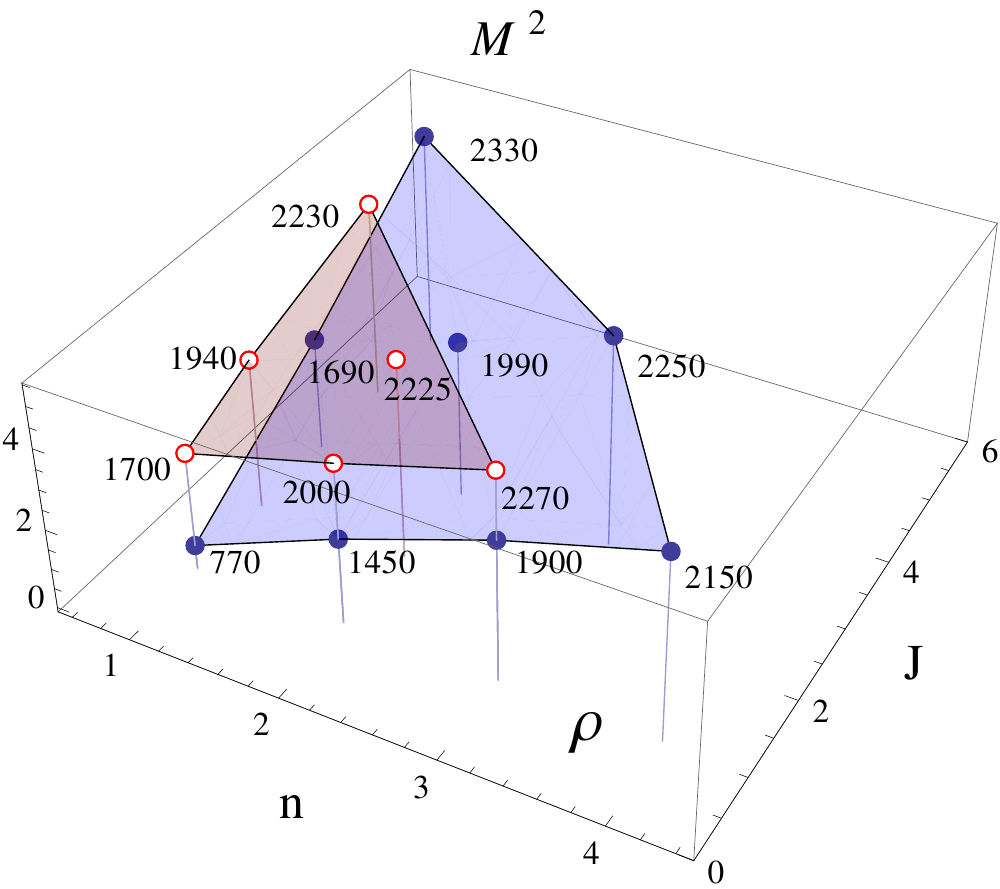}
  \caption{(color online) The $(n,J,M^2)$ Regge planes for the $\eta$ family and the
$\rho$ family. \label{nJM2plane}}
\end{figure*}

\section{Joint $(n,J,M^2)$ fits \label{sec:merge}}

A summary of the estimated radial and angular slopes is given
in Fig.~\ref{muJall}.
We consider a weighted average estimate for a common (universal) slope trajectory,
including the studied trajectories both in the $(n,M^2)$ and the
$(J,M^2)$ planes. This yields
$\mu^2=\beta=1.26(3)$~GeV$^2$. When the spread of central
values for all the trajectories is used, we obtain
$\mu^2=\beta=1.27(14)$~GeV$^2$. The trend to produce a number closer to the
radial slope reflects a larger sample. As a matter of fact, when  we
weight both $n$ and $J$ trajectories equally, we get
$\mu^2=\beta=1.26(3)$~GeV$^2$.

In the previous sections we have carried out the analysis for the $(n,M^2)$ and $(J,M^2)$ planes,
with the indication that $\mu^2 > \beta$ at a significant statistical
level (3.4 standard deviations).  This strongly suggests a careful
reconsideration of the findings of Ref.~\cite{Afonin:2006wt}, where a
common fit of Eq.~(\ref{eq:afo}) for the $(n,M^2)$ and $(J,M^2)$
planes was proposed. Technically, the joint analysis presented in this
Section is different form the separate analyses of
Sect.~\ref{sec:radial-Regge} and \ref{sec:J-Regge} in the following
important detail. In the separate fits the constants $M_0^2$ and
${M'}_0^2$ in the formulas $M^2(n)=\mu^2 n + M_0^2$ and $M^2(n)=\beta
J + {M'}_0^2$ were treated as unrelated parameters, even in the same
family of states.  On the contrary, formula (\ref{eq:afo}), with a
common parameter $c$ for a given family, relates the ``offset''
constants $M_0^2$ and ${M'}_0^2$, providing a constraint to the
statistical analysis.

Therefore, to look closer at the issue of universality of the radial
and angular-momentum slopes, we analyze each sector independently with
two different plane fit functions: the non-universal formula
\begin{eqnarray}
M^2=a n + b J + c \label{eq:nonu}
\end{eqnarray}
on one hand, and the universal formula
\begin{eqnarray}
M^2=a(n+J)+c \label{eq:u}
\end{eqnarray}
on the other hand. With the $c$ parameter fixed for the whole family,
the states with different values of $n$ and $J$ belong to a {\em Regge
  plane}. This is illustrated in Fig.~\ref{nJM2plane} for the $\rho$-
and $\eta$-families as an example.

Our numerical results are collected  in
Table~\ref{TablePlanes}. Several meson sectors can be placed on two almost
parallel planes. The nomenclature used is as follows: after the name of
each family, the subindex quoted refers to the states with the
particular angular quantum number used on the plane. For example,
$\rho_{135}$ means the set of all the $\rho$ states with angular-momentum $J=1,3,5$, and with all the possible radial quantum numbers.

The numbers in Table~\ref{TablePlanes} show a few interesting
features. The most important one is that for each family of states the
fit with Eq.~(\ref{eq:nonu}) is preferred over the fit with Eq.~(\ref{eq:u}) (judging by the
different $\chi^2/{\rm DOF}$ values). Moreover, we generically find $a>b$.
The offset parameter $c$ also seems to be stable through all
the planes, although less stable than the radial and angular slope
parameters.

The fact that the ratio of the radial to angular slope, denoted as
$R$, grows with the quark mass may be a generic and physically
relevant feature. Note that for the heavy quarkonia the joint fit is
compatible with the formula $M^2 = a(2n+J)+c$
\cite{Gershtein:2006ng}. Thus, it may be that $R$ is close to unity
for light mesons, equal 2 for heavy mesons, and assumes an
intermediate value for hidden-strangeness states.

From Table~\ref{TablePlanes} it is worth stressing how close the
states from the $\rho$ and the $\omega$ families are.  For $a_{02}$
plane, we have assigned to $a_2(2030)$ and $a_2 (2255)$ radial quantum
numbers $n=2$ and $3$, respectively.  Otherwise (with $n=1$ and $2$),
the angular slope would be larger than the radial slope. The
$h_{13}$, $b_{13}$, and $f_{13}$ have systematically smaller $a$ and
$b$ parameters from the remaining families, although very similar
among themselves. As commented already, this fact may be caused by the lack
of states in these sectors.  States with higher angular-momentum (when
discovered) would lead to better and more reliable determination of
the Regge plane parameters.  Similar comments apply to the $\phi_{13}$
plane.

\begin{table*}
  \centering
\renewcommand{\arraystretch}{1.5}
{
\begin{tabular}{|c||c|c|c|c||c|c|c|}
\hline
  &  \multicolumn{4}{c|}{$M^2=a n+bJ+c$} & \multicolumn{3}{c|}{$M^2=a(n+J)+c$}  \\
\hline
 & a & b & c & $\chi^2/{\rm DOF}$ & a & c & $\chi^2/{\rm DOF}$ \\
\hline
$a_{13}$       & 1.41(45) & 1.11(32) & -1(1)     &  0.09 & 1.20(28) & -0.89(94) &  0.19 \\
$\eta_{024}$   & 1.36(5)  & 1.21(9)  & -1.06(5)  &  0.22 & 1.33(4)  & -1.03(4)  &  0.51 \\
$\eta_{02_{ss}}$& 1.36(14) & 1.27(22) & -0.50(28) &  0.44 & 1.34(13) & -0.47(27) &  0.35 \\
$\rho_{135}$   & 1.36(12) & 1.12(9)  & -1.87(23) &  0.40 & 1.21(7)  & -1.79(22) &  0.73 \\
$\pi_{024}$    & 1.47(10) & 1.27(10) & -1.45(10) &  0.29 & 1.36(6)  & -1.34(6)  &  0.50 \\
$a_{246}$      & 1.35(25) & 1.06(16) & -1.75(48) &  0.14 & 1.15(13) & -1.70(48) &  0.33 \\
$a_{02}$     & 1.35(24) & 0.78(24) & -0.39(27) &  0.53 & 1.06(9) & -0.09(13) &  0.90 \\
$f_{0246}$     & 1.38(13) & 0.64(8)  & 0.04(33) &  0.85 & 0.76(8)  & 1.06(29)  &  5.03 \\
$f_{02}$       & 1.34(11) & 0.69(6) & -0.38(15) &  0.14 & 0.84(6) & 0.13(11) &  5.66 \\
$\omega_{135}$ & 1.42(11) & 0.98(8)  & -1.78(11) &  0.63 & 1.16(5)  & -1.70(10) &  1.77 \\
$h_{13}$       & 1.17(23) & 0.75(19) & -0.53(62) &  0.02 & 0.93(13) & -0.37(61) &  0.46 \\
$b_{13}$       & 1.15(17) & 0.72(15) & -0.35(32) &  0.05 & 0.91(9)  & -0.29(32) &  0.88 \\
$f_{13}$       & 1.19(15) & 0.70(19) & -0.24(17) &  0.07 & 0.98(8)  & -0.33(17) &  0.95 \\
$\phi_{13}$    & 1.84(6)  & 1.20(08) & -2.0(1)   &  0.06 & 1.59(5)  & -2.15(10) &  19.5 \\
\hline
\end{tabular}
}
\caption{Regge-plane fits combining both radial and angular-momentum
  trajectories (see main text for details). \label{TablePlanes}}
\end{table*}

Considering only the planes with no large hidden strangeness content and with six or more states (excluding then the $h_{13}$, $b_{13}$, $f_{13}$, and $\phi_{13}$ planes)\footnote{The $f_{02}$ is not consider either due to the arbitrariness on the selection of its components.} we
obtain our global fit with the result
\begin{equation} 
M^2= 1.38(4)n + 1.12(4) J -1.25(4)\, . \label{eq:finfit}
\end{equation}
Therefore, the $a=\mu^2$ parameter reads $a=1.38(4)$~GeV$^2$ for the
global Regge-plane fit, compatible with Eq.~(\ref{eq:mu_g}).  The
$b=\beta$ parameter reads $b=1.12(4)$~GeV$^2$, also close to the value
of Eq.~(\ref{eq:beta_g}).  We note that
$a > b$ at the level of $4.5$ standard deviations. In this estimate
we take the geometric average of the individual errors (equal $0.04$) for the standard deviation of the
difference $a-b$. Therefore the joint analysis points at a lack of universality of the Regge slopes.

The right part of Table~\ref{TablePlanes} shows the result of the fit,
where universality is imposed. This fit cannot be statistically rejected based on
the values of $\chi^2$, however, it is somewhat worse than without the
universality constraint.

\section{Conclusions \label{sec:concl}}

In this paper we have reanalyzed, with the help of the up-to-date
PDG tables~\cite{Nakamura:2010zzi}, the linear radial and angular-momentum
Regge trajectories considered in
Ref.~\cite{Anisovich:2000kxa}, including in the fits the width of
each state as an estimate of the error of the resonance mass (the
half-width rule). As we have explained this is a reasonable way to
smooth out resonance profile information, which makes the very
definition of the resonance mass ambiguous. Moreover, this choice allows to
undertake an error analysis, not carried out in
Ref.~\cite{Anisovich:2000kxa}. Furthermore, we have argued that such a
procedure fully complies to the large-$N_c$ viewpoint and actually
suggests an interesting interpretation: the Regge-fitted masses are
considered to be the leading-$N_c$ contribution to the mass of the
resonance. This incorporates a desirable flexibility as to what should
the Regge fit be compared to. The squared mass of each meson is
then represented as $M_n^2 = M^2 \pm \Gamma M$, where $M$ is its mass
and $\Gamma$ is its width.

Generally, we reproduce the results of Ref.~\cite{Anisovich:2000kxa}
when no uncertainties are included. This only reflects the robustness
of the main features of the PDG compilation along the last 10 years, although some
numerical values of the masses have changed and, furthermore, some new
states, partly predicted by the pioneering radial Regge analysis of
Ref.~\cite{Anisovich:2000kxa}, have been added. From our results it follows that
there is no need to consider further new states to get an acceptable Regge
description. This is consistent with an assertion of a complete mesonic spectrum up to
the highest energies considered in our work.

We have also addressed the issue of the {\em universality} of
radial and angular-momentum slopes within the errors deduced from the
linear regression analysis with weights provided with the half-width rule.
Our joint analysis in the $(n,J,M^2)$ Regge planes
indicates, at a statistically significant level of 4.5 standard deviations, that the radial slope
is larger from the angular-momentum slope. Thus no strict universality
of slopes occurs in the light non-strange meson spectra.

\appendix

\section{Dependence on resonance profile}
\label{sec:app}

In this appendix we show the independence of our results on the shape of the resonance profiles, and
hence support our $\chi^2$-statistical treatment. The $\chi^2$-nature
of the fit relies implicitly on the assumption that the probability of
having a resonance with mass $M$ and width $\Gamma$ is of a Gaussian
shape,
\begin{eqnarray}
P(\sqrt{s}) = C e^{-\frac{(\sqrt{s}-M)^2}{\Gamma M}}  \, ,
\label{eq:gauss1}
\end{eqnarray}
with $C$ a normalization constant, whereas for the squared mass one has
\begin{eqnarray}
P(s) = C' e^{-\frac{(s-M^2)^2}{2 \Gamma^2  M^2}} \, .
\label{eq:gauss2}
\end{eqnarray}
The $\chi^2$-fit then corresponds to applying the
maximum-likelihood-method (MLM) and maximize with respect to $a$ and
$M_0$ the function
\begin{eqnarray}
L( a, M_0 ; \{ \Gamma_n, M_n \} ) =\prod_{n=1}^N P(s_n, \Gamma_n, M_n) \, ,
\label{eq:MLM}
\end{eqnarray}
where $s_n= a n + M_0^2$. This is the way the half-width rule is
implemented in practice, i.e., by assuming short tails in the
mass-distribution.  On the other hand, from analyticity arguments the
resonance profile function should be
of a Breit-Wigner form, at
least for sufficiently narrow resonances. 
Let us consider for definiteness the parameterization
of a complex resonance propagator at a given CM energy squared, $s$,
\begin{eqnarray}
D(s)=\frac{1}{s-M^2-i \Gamma \sqrt{s}} \, .
\end{eqnarray}
The $\sqrt{s}$ in the denominator ensures that we have a pole on the
second Riemann sheet (we neglect threshold effects). Likewise, we also
have a pure imaginary amplitude at the real resonance value
$s=M^2$. The probability for such a mass distribution corresponds to
the imaginary part~\footnote{We are appealing to the
  Lehman representation for a resonance as obtained from a CM-energy
  dispersion relation of the scattering process, see, e.g.,
  Ref.~\cite{CalleCordon:2009ps} for a discussion in the context of the
  $\pi\pi$-scattering.},  namely,
\begin{eqnarray}
P_{\rm BW}(s)= Z \frac{\Gamma \sqrt{s}}{(s-M^2)^2 +\Gamma^2 s)} \, ,
\label{eq:BWdist}
\end{eqnarray}
where $Z$ is a suitable normalization constant. Thus, we may apply the
MLM to Eq.~(\ref{eq:MLM}) for N resonances fulfilling the Regge
formula and maximize with respect to $a$ and $M_0$.

\begin{figure}
\centering
  \includegraphics[width=0.45\textwidth]{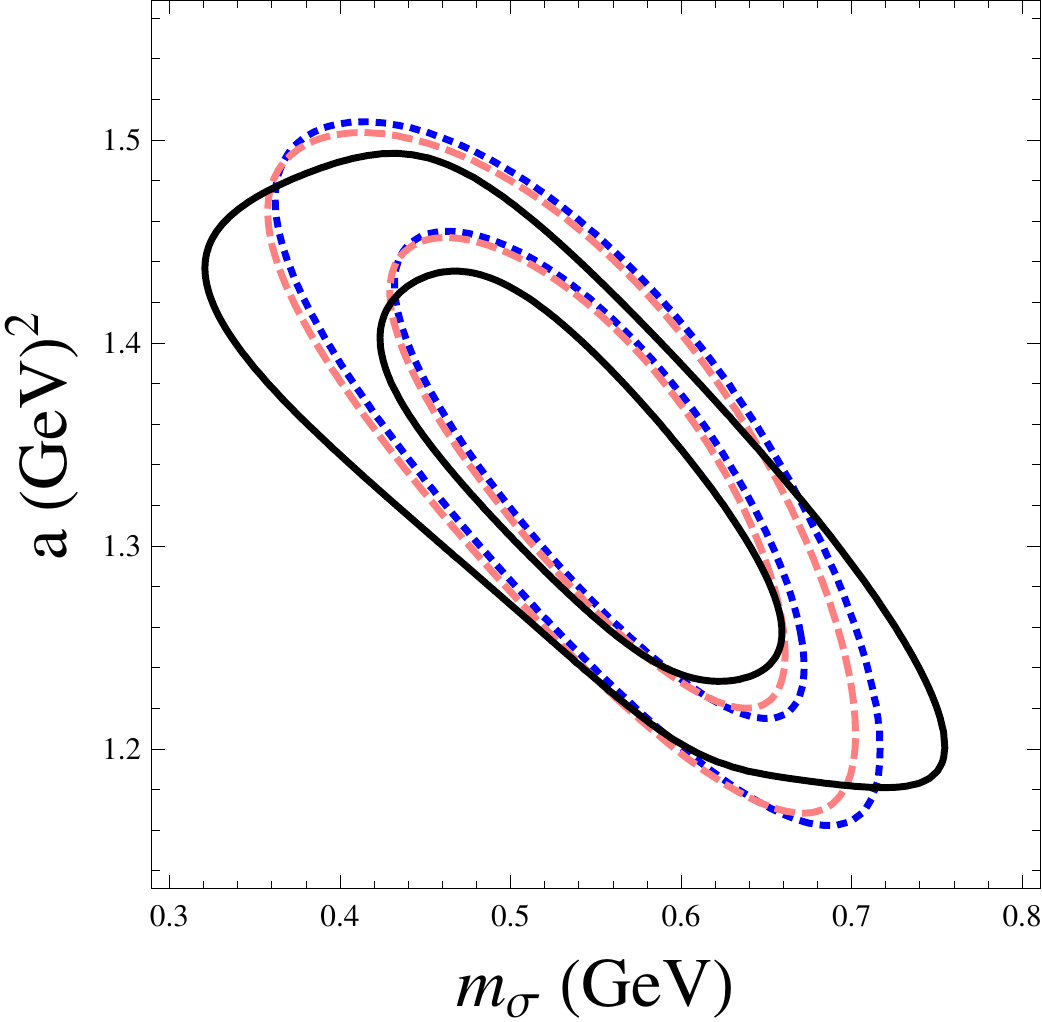}
  \caption{(color online) The $68\%$ and $95\%$ relative
confidence-level contours of the radial Regge trajectories
for all scalars in the PDG assuming several resonance profiles. 1) Gaussian
at the level of the mass (dashed line), 2) Gaussian
at the level of the squared-mass (dotted line) and 3) Breit-Wigner shape (black.solid).}\label{res-profs}
\end{figure}

As an specific example, to illustrate the difference between the
Gaussian, Eq.~(\ref{eq:gauss2}), and the Breit-Wigner,
Eq.~(\ref{eq:BWdist}), profiles to the set of all $0^{++}$ scalars
listed in the PDG (see also Fig.~\ref{f0plot}) as discussed in
Refs.~\cite{RuizArriola:2010fj,Arriola:2011en}, where a joint formula
for the trajectories,
\begin{eqnarray}
s_n = \frac{a}{2} n + m_\sigma^2,
\end{eqnarray}
was proposed. Maximizing $P( a, m_\sigma ; \{ M_n, \Gamma_n \})
$ with respect to $a$ and $m_\sigma$ and using Eqs.~(\ref{eq:gauss1}),
(\ref{eq:gauss2}), and (\ref{eq:BWdist}) yields the most likely values
$m_\sigma = 0.545$, $0.557$, $0.562~{\rm GeV}$, and $a = 1.330$, $1.336$,
$1.334~{\rm GeV}^2$, respectively. The Gaussian cases correspond to the
$\chi^2$-analysis of Refs.~\cite{RuizArriola:2010fj,Arriola:2011en}.
Similarly to the $\chi^2$-method, the errors can be determined by looking
at the locus of the relative probability $P (a, m_\sigma)/P_{\rm max} \le
(e^{-\Delta \chi^2/2})$, which for two variables yields $\Delta \chi^2
= 2.3$ and $4.7$ for the $68 \%$ and $95 \%$ confidence levels, respectively. We
show the results in Fig.~\ref{res-profs} where, as we can see, the
resonance shape does not play a role.

Of course, one may object to the previous confidence level analysis
that for non-Gaussian probabilities mode (most-likely) and mean
(average) are different. Indeed, the application of
\begin{eqnarray}
\langle A \rangle = \int da \int d m_\sigma A (a, m_\sigma) L_{\rm BW}( a, m_\sigma ; \{ M_n,\Gamma_n \})
\end{eqnarray}
 yields $ \langle a \rangle = 1.34~{\rm GeV}^2$  and
 $\langle m_\sigma \rangle= 0.53 {\rm GeV}$ for the mean values, whereas the mode is at
 $m_\sigma =0.562~{\rm GeV}$ and $a = 1.334~{\rm GeV}^2$.
 Thus, for the Breit-Wigner case the mean and the mode are different and the errors
 are not defined by standard confidence level rules with $\langle
 A \rangle \pm \sqrt{\langle A^2 \rangle - \langle A \rangle^2} $, for
 which we get $ a = 1.34(8) {\rm GeV}^2$, $m_\sigma = 0.53(11) {\rm
   GeV}$ with a correlation $r(a,m_\sigma)=-0.77$.
 A way to sort this out is to define the equal probability contours, to
 integrate inside the inner region for a given confidence level,
\begin{eqnarray}
p(z) = \int da d m_\sigma P (a,m_\sigma) \Theta( P (a,m_\sigma)-z) \, ,
\end{eqnarray}
and to search for a $z$ such that $p(z)= 0.68$. The resulting contour
resembles strongly Fig.~\ref{res-profs}, reinforcing the conclusion
that the shape of the resonance profile is irrelevant
for the analyses of this work.



\end{document}